\documentclass[a4paper,11pt]{article}
\usepackage{pos}
\usepackage{microtype}
\usepackage{graphicx}
\usepackage{xcolor}
\usepackage{palatino}
\usepackage{amsmath}
\usepackage{tikz}
\usepackage{tikz-feynman}
\usepackage{amsfonts}
\usepackage{dsfont}
\usepackage{slashed}
\usepackage{braket}
\usepackage{amsthm}
\usepackage{pgfplots}

\newcommand{\boundary}{\partial}

\newcommand{\vertices}{\mathcal{V}}

\newcommand{\threerep}{f^{\text{3d}}}

\tikzset{
    ->-/.style={decoration={
  markings,
  mark=at position .5 with {\arrow{>}}},postaction={decorate}},
    -<-/.style={decoration={
  markings,
  mark=at position .5 with {\arrow{<}}},postaction={decorate}},
}

\title{Derivation of the Cross-Free Family representation for the box diagram}
\author[a]{Zeno Capatti}
\affiliation[a]{Institute for Theoretical Physics,
  University of Bern, \\ Sidlerstrasse 5, 3012, Bern \\ Switzerland}

\emailAdd{zeno.ca@gmail.com}

\abstract{I work out in full detail the derivation of the Cross-Free Family (CFF) representation for the box diagram, and highlight the differences with its Time Ordered Perturbation Theory (TOPT) representation. I briefly discuss implications for the threshold singularity structure of the diagram.}

\FullConference{%
  RADCOR 2023 – 16th International Symposium on Radiative Corrections: Applications of Quantum Field Theory to Phenomenology, \\
  Sunday 28th May 2023 - Friday 2nd June 2023,\\
  Crieff, Scotland. 
}

\begin{document}
\maketitle

\section{Introduction}

\noindent This document contains an explicit example of the derivation of the Cross-Free Family representation hypothesized in ref.~\cite{Capatti:2020xjc} and then derived systematically in ref.~\cite{Capatti:2022mly}. It is a three-dimensional representation~\cite{Aguilera_Verdugo_2020,Aguilera_Verdugo_2021,Bierenbaum_2010,bobadilla2021lotty,Borinsky:2022msp,Capatti_2019,Capatti:2022mly,capatti2020manifestly,Catani_2008,Kromin:2022txz,Ramirez-Uribe:2020hes,Ramirez-Uribe:2022sja,Runkel_2019,Runkel_2020,Sborlini_2021} in which the threshold singularity structure of the Feynman diagram becomes especially constrained and in which spurious singularities are absent. These constraints are related to the graph-theoretic notions of connectedness and crossing that recur often in analyses of the singularity structure of Feynman diagrams~\cite{Abreu:2014cla,Abreu:2017ptx,Benincasa:2021qcb,Berghoff:2020bug,bourjaily2020sequential,Gardi:2022khw,Hannesdottir:2022bmo,Kreimer:2020mwn,Kreimer:2021jni,Sborlini_2021}, and will be derived by using tools from graph-theory and convex geometry (specifically, by using properties of Fourier transforms of polytopes~\cite{polytope_fourier}). In~\cite{Sterman:2023xdj}, a novel and interesting derivation of the representation based on posets has been proposed.

\noindent We are interested in performing analytically the following integral 
\begin{equation}
\threerep_\Box=\int \frac{\mathrm{d}k^0 }{2\pi}\frac{\mathcal{N}(\{k^0+p_{1i}^0\}_{i=1}^4;\{\vec{k}+\vec{p}_{1i}\}_{i=1}^4)}{((k+p_{11})^2-m_1^2) ((k+p_{12})^2-m_2^2) ((k+p_{13})^2-m_3^2) ((k+p_{14})^2-m_4^2)},
\end{equation}
with $p_{1i}=\sum_{j=1}î p_i$. The full Feynman diagram for the box diagram is written in terms of $\threerep_\Box$ as follows:
\begin{equation}
\label{eq:orig_routing}
\resizebox{3.5cm}{!}{\raisebox{-1.7cm}{
\begin{tikzpicture}

    \begin{feynman}
        \vertex(1);

        \vertex[right = 2cm of 1](2);

        \vertex[below = 2cm of 2](3);

        \vertex[left = 2cm of 3](4);
        
         \diagram*[large]{	
            (1) -- [-,line width=0.6mm,momentum=\(k+p_1\)] (2),
            (2) -- [-,line width=0.6mm,momentum=\(k+p_1+p_2\)] (3),
            (3) -- [-,line width=0.6mm,momentum=\(k-p_4\)] (4),
            (4) -- [-,line width=0.6mm,momentum=\(k\)] (1),
        }; 
      \path[draw=black, fill=black] (1) circle[radius=0.05];
       \path[draw=black, fill=black] (2) circle[radius=0.05];
       \path[draw=black, fill=black] (3) circle[radius=0.05];
       \path[draw=black, fill=black] (4) circle[radius=0.05];
       
    \end{feynman}
    
\end{tikzpicture}
}}=\int \frac{\mathrm{d}^3\vec{k} }{(2\pi)^3}\threerep_\Box .
\end{equation}
The Feynman diagram corresponds to a graph which is characterised by a set of vertices, $\mathcal{V}=\{v_1,v_2,v_3,v_4\}$, and edges, $\mathcal{E}=\{e_1,e_2,e_3,e_4\}$, with labels assigned according to the following convention
\begin{equation}
\label{eq:box_labelling}
\resizebox{2.5cm}{!}{\raisebox{-0.5cm}{
\begin{tikzpicture}

    \begin{feynman}
        \vertex(1);

        \vertex[right = 2cm of 1](2);

        \vertex[below = 2cm of 2](3);

        \vertex[left = 2cm of 3](4);

        \vertex[above right =0.2cm and 0.8cm of 1](L1) {\resizebox{0.5cm}{!}{$e_2$}};
        \vertex[below right =0.2cm and 0.8cm of 4](L2) {\resizebox{0.5cm}{!}{$e_4$}};

        \vertex[below right =0.8cm and 0.2cm of 2](L3) {\resizebox{0.5cm}{!}{$e_3$}};
        \vertex[below left =0.8cm and 0.2cm of 1](L4) {\resizebox{0.5cm}{!}{$e_1$}};

        \vertex[above left =0.05cm and 0.05cm of 1](LV1) {\resizebox{0.5cm}{!}{$v_1$}};
        \vertex[above right =0.05cm and 0.05cm of 2](LV2) {\resizebox{0.5cm}{!}{$v_2$}};
        \vertex[below right =0.05cm and 0.05cm of 3](LV3) {\resizebox{0.5cm}{!}{$v_3$}};
        \vertex[below left =0.05cm and 0.05cm of 4](LV4) {\resizebox{0.5cm}{!}{$v_4$}};

         \diagram*[large]{	
            (1) -- [-,line width=0.6mm] (2),
            (2) -- [-,line width=0.6mm] (3),
            (3) -- [-,line width=0.6mm] (4),
            (4) -- [-,line width=0.6mm] (1),
        }; 
      \path[draw=black, fill=black] (1) circle[radius=0.05];
       \path[draw=black, fill=black] (2) circle[radius=0.05];
       \path[draw=black, fill=black] (3) circle[radius=0.05];
       \path[draw=black, fill=black] (4) circle[radius=0.05];
       
    \end{feynman}
    
\end{tikzpicture}
}}.
\end{equation}
 We are ready to start deriving the Cross-Free Family representation.

\section{Acyclic graphs}

\noindent We introduce one auxiliary integration variable for each edge, named $q_i^0$, which simply equals the energy flowing through that edge, expressed as a linear combination of the energy loop variable $k^0$ and the energies of the external momenta
\begin{equation}
\threerep_\Box=\int \frac{\mathrm{d}k^0 }{2\pi} \int \left[\prod_{i=1}^4 \frac{\mathrm{d}q_i^0}{(q_i^0)^2-E_i^2}\delta(q_i^0-k^0-p_{1i}^0)\right]\mathcal{N}(\{q_i^0\}_{i=1}^4;\{\vec{k}+\vec{p}_{1i}\}_{i=1}^4),%\frac{\delta(q_1^0-k^0)}{(q_1^0)^2-E_1^2}\frac{}{}\frac{\delta(q_1^0-k^0-p_{12}^0)}{(q_3^0)^2-E_3^2}\frac{\delta(q_1^0-k^0-p_{13}^0)}{(q_4^0)^2-E_4^2},
\end{equation}
having introduced the on-shell energies
\begin{equation}
E_i=\sqrt{|\vec{k}+\vec{p}_{1i}|^2+m_i^2-i\varepsilon}.
\end{equation}
We also suppressed the dependence on the spatial components of the momenta in the numerator, for the sake of compactness. Using the Fourier representation of the Dirac delta function, we obtain
\begin{equation}
\threerep_\Box=\int \frac{\mathrm{d}k^0 }{2\pi} \int \left[\prod_{i=1}^4 \frac{\mathrm{d}q_i^0\mathrm{d}\tau_i}{2\pi}\frac{\exp\left\{i\tau_i\left(q_i^0-k^0-p_{1i}^0\right)\right\}}{(q_i^0)^2-E_i^2}\right]\mathcal{N}(\{q_i^0\}_{i=1}^4), \nonumber
\end{equation}
which can be re-written in order to highlight the simplicity of the $q_i^0$ integration
\begin{equation}
\label{eq:cff_box_pre_residue}
\threerep_\Box=\int \frac{\mathrm{d}k^0 }{2\pi} \int \left[\prod_{i=1}^4\mathrm{d}\tau_ie^{i\tau_i\left(-k^0-p_{1i}^0\right)}\right]\int \left[\prod_{j=1}^4 \frac{\mathrm{d}q_j^0}{2\pi} \frac{e^{i q_j^0 \tau_j}}{(q_j^0)^2-E_j^2}\right]\mathcal{N}(\{q_i^0\}_{i=1}^4).
\end{equation}
In particular, the advantage of introducing the auxiliary integration variables is now evident: energy conservation is resolved through the oscillatory behaviour of the complex exponentials and the integration over the energies $q_i^0$ of each propagator is trivial to perform. In particular, the following identity holds:
\begin{align}
\label{eq:cff_box_energy_integral}
 \int \left[\prod_{j=1}^4 \frac{\mathrm{d}q_j^0}{2\pi} \frac{e^{i q_j^0 \tau_j}}{(q_j^0)^2-E_j^2}\right]\mathcal{N}(\{q_i^0\}_{i=1}^4)=
 \sum_{\vec{\sigma}\in\{\pm\}^4} \mathcal{N}(\{\sigma_iE_i\}_{i=1}^4)\prod_{j=1}^4\frac{\Theta(-\sigma_j \tau_j) e^{i \sigma_j E_j \tau_j}}{2 i E_j},
\end{align}
obtained by straight-forward contour integration. Notably, such an identity holds only if the residue at infinity of the integrand, in each of the $q_j^0$ variables, vanishes. This in turn imposes a constraint on the shape of the numerator, namely that it can be linear at most in each of the $q_i^0$ variables. In other words, it must take the form
\begin{equation}
\mathcal{N}(\{q_i^0\}_{i=1}^4)=\sum_{i_1=0}^1\sum_{i_2=0}^1\sum_{i_3=0}^1\sum_{i_4=0}^1 a_{i_1i_2i_3i_4}(q_1^0)^{i_1}(q_2^0)^{i_2}(q_3^0)^{i_3}(q_4^0)^{i_4}. 
\end{equation}
Substituting eq.~\eqref{eq:cff_box_energy_integral} in eq.~\eqref{eq:cff_box_pre_residue}, and performing the change of variables $\tau_j\rightarrow \sigma_j\tau_j$, we obtain
\begin{equation}
\threerep_\Box=\sum_{\vec{\sigma}\in\{\pm\}^4}\frac{\mathcal{N}(\{\sigma_n E_n\}_{n=1}^4)}{\prod_{m=1}^4 2 i E_m}\int \frac{\mathrm{d}k^0 }{2\pi} \int \left[\prod_{i=1}^4\mathrm{d}\tau_i\Theta(\tau_i) e^{-i E_i \tau_i}\right]\left[\prod_{j=1}^4 e^{-i\sigma_j\tau_j(k^0+p_{1j}^0)}\right],
\end{equation}
having used that $\sigma_j^2=1$. Finally, integration in $k^0$ is trivial by using the Fourier representation of the Dirac delta to rewrite the integration in $k^0$. In particular
\begin{equation}
\threerep_\Box=\sum_{\vec{\sigma}\in\{\pm\}^4} \frac{\mathcal{N}(\{\sigma_n E_n\}_{n=1}^4)}{\prod_{m=1}^4 2 i E_m} \hat{\mathds{1}}_{\vec{\sigma}} , \quad  \hat{\mathds{1}}_{\vec{\sigma}} =\int \left[\prod_{i=1}^4\mathrm{d}\tau_i \Theta(\tau_i) e^{-i (E_i-\sigma_i p_{1i}^0) \tau_i}\right]\delta\left(\sum_{j=1}^4\sigma_j\tau_j\right).
\end{equation}
The sum over vector signs $\vec{\sigma}$ can be interpreted as a sum over orientations of the edges of the box diagram. The sign vector $\vec{\sigma}=(1,1,1,1)$, which we represent by assigning an arrow to the edge, corresponds to the orientation of the edges chosen for the original routing of eq.~\eqref{eq:orig_routing}:
\begin{equation}
\label{eq:orig_routing_2}
G_{\vec{\sigma}_1}=\resizebox{1.5cm}{!}{\raisebox{-1.cm}{
\begin{tikzpicture}

    \begin{feynman}
        \vertex(1);

        \vertex[right = 2cm of 1](2);

        \vertex[below = 2cm of 2](3);

        \vertex[left = 2cm of 3](4);
        
         \diagram*[large]{	
            (1) -- [->-,line width=0.6mm] (2),
            (2) -- [->-,line width=0.6mm] (3),
            (3) -- [->-,line width=0.6mm] (4),
            (4) -- [->-,line width=0.6mm] (1),
        }; 
      \path[draw=black, fill=black] (1) circle[radius=0.05];
       \path[draw=black, fill=black] (2) circle[radius=0.05];
       \path[draw=black, fill=black] (3) circle[radius=0.05];
       \path[draw=black, fill=black] (4) circle[radius=0.05];
       
    \end{feynman}
    
\end{tikzpicture}
}}.
\end{equation}
The sign vector $\vec{\sigma}_2=(-1,-1,-1,-1)$ corresponds to the orientation obtained from that of eq.~\eqref{eq:orig_routing_2} by flipping the arrows of all edges. The sign vector $\vec{\sigma}_3=(-1,1,-1,1)$, instead, corresponds to the orientation obtained from that of eq.~\eqref{eq:orig_routing_2} by flipping the arrow for the edges with momentum $q_1$ and $q_3$:
\begin{equation}
\label{eq:graph_or_5}
G_{\vec{\sigma}_3}=\resizebox{1.5cm}{!}{\raisebox{-1.cm}{
\begin{tikzpicture}

    \begin{feynman}
        \vertex(1);

        \vertex[right = 2cm of 1](2);

        \vertex[below = 2cm of 2](3);

        \vertex[left = 2cm of 3](4);
        
         \diagram*[large]{	
            (1) -- [->-,line width=0.6mm] (2),
            (2) -- [-<-,line width=0.6mm] (3),
            (3) -- [->-,line width=0.6mm] (4),
            (4) -- [-<-,line width=0.6mm)] (1),
        }; 
      \path[draw=black, fill=black] (1) circle[radius=0.05];
       \path[draw=black, fill=black] (2) circle[radius=0.05];
       \path[draw=black, fill=black] (3) circle[radius=0.05];
       \path[draw=black, fill=black] (4) circle[radius=0.05];
       
    \end{feynman}
    
\end{tikzpicture}
}}.
\end{equation}
In general, we can associate to $\vec{\sigma}$ the directed graph $G_{\vec{\sigma}}$ constructed in the way represented above. To conclude this section, we will now look more closely at the sum over vectors $\vec{\sigma}$ and show that two terms of this sum vanish identically. They correspond to the two vectors $\sigma_1=(1,1,1,1)$ and $\sigma_2=(-1,-1,-1,-1)$. Indeed,  
\begin{equation}
\hat{\mathds{1}}_{\vec{\sigma}_1}=\int \left[\prod_{i=1}^4\mathrm{d}\tau_i \Theta(\tau_i) e^{-i (E_i-p_{1i}^0) \tau_i}\right]\delta\left(\tau_1+\tau_2+\tau_3+\tau_4\right)=0
\end{equation}
since the constraints imposed by the Heaviside theta functions, namely that $\tau_i>0$, cannot be satisfied simultaneously with the linear constraint, imposed by the Dirac delta function, that $\tau_1+\tau_2+\tau_3+\tau_4=0$. Looking at the directed graphs $G_{\vec{\sigma}_1}$ and $G_{\vec{\sigma}_2}$, we realise that they are all and only the directed graphs that have a directed cycle within them. In other words, we can walk starting from one of their vertices and, following the direction of the arrows, return to the original vertex. The graphs that escape this analysis are precisely those that do not have a directed cycle within them; in technical jargon, they are called \emph{acyclic graphs}. In turn, we can write
\begin{equation}
\threerep_\Box=\sum_{\substack{\vec{\sigma} \text{ s.t.} \\ G_{\vec{\sigma}} \text{ is acyclic}}} \frac{\mathcal{N}(\sigma_1E_1,\sigma_2E_2,\sigma_3E_3,\sigma_4E_4)}{\prod_{j=1}^4 2i E_j} \hat{\mathds{1}}_{\vec{\sigma}}.
\end{equation}
This result is dual to that obtained within Flow-Oriented Perturbation Theory~\cite{Borinsky:2022msp}. We now briefly expand on the mathematical interpretation of the object $\hat{\mathds{1}}_{\vec{\sigma}}$.

\section{Fourier transform of cones}

\noindent The Heaviside Theta functions and the Dirac delta distribution can be absorbed in a redefinition of the integration domain. More specifically, we collect the relevant integration variables in a vector $\boldsymbol{\tau}=(\tau_1,\tau_2,\tau_3,\tau_4)$, the corresponding on-shell energies in another $\mathbf{E}=(E_1,E_2,E_3,E_4)$, and finally the energy shifts of the edges in yet another vector $\mathbf{p}^0_{\vec{\sigma}}=(\sigma_1 p_{11}^0,\sigma_2 p_{12}^0,\sigma_3 p_{13}^0,\sigma_4 p_{14}^0)$. We then define the integration domain
\begin{equation}
\mathcal{K}_{\vec{\sigma}}=\left\{\boldsymbol{\tau}\in\mathbb{R}_+^4 \ \bigg| \ \sigma_1\tau_1+\sigma_2\tau_2+\sigma_3\tau_3+\sigma_4\tau_4=0 \right\}.
\end{equation}
$\mathcal{K}_{\vec{\sigma}}$ is a convex cone, as it is the intersection of the positive orthant (a convex cone) with a hyperplane. In terms of the newly-defined vectors and the cone $\mathcal{K}_{\vec{\sigma}}$, we may rewrite
\begin{equation}
\hat{\mathds{1}}_{\vec{\sigma}}=\int_{\mathcal{K}_{\vec{\sigma}}} \mathrm{d}\boldsymbol{\tau} e^{-i\boldsymbol{\tau}\cdot (\mathbf{E}-\mathbf{p}^0_{\vec{\sigma}})}=\int_{\mathbb{R}^4} \mathrm{d}\boldsymbol{\tau} e^{-i\boldsymbol{\tau}\cdot (\mathbf{E}-\mathbf{p}^0_{\vec{\sigma}})} \mathds{1}_{\vec{\sigma}}(\boldsymbol{\tau}),
\end{equation}
i.e. the Fourier transform of the characteristic function $\mathds{1}_{\vec{\sigma}}$ of the set $\mathcal{K}_{\vec{\sigma}}$, defined as usual:
\begin{equation}
\mathds{1}_{\mathcal{K}_{\vec{\sigma}}}(\boldsymbol{\tau})=\begin{cases}
1 \quad &\text{if } \boldsymbol{\tau} \in \mathcal{K}_{\vec{\sigma}} \\
0 \quad &\text{otherwise}
\end{cases}.
\end{equation}
Following the results of the previous section, we conclude that $\mathcal{K}_{\vec{\sigma}}$ is the empty set if $G_{\vec{\sigma}}$ is not an acyclic graph. Computing the Fourier transform $\hat{\mathds{1}}_{\vec{\sigma}}$ requires to find a triangulation of the cone $\mathcal{K}_{\vec{\sigma}}$.

\section{Triangulations and the edge-contraction operation}

\subsection{An example}

\noindent For the purposes of this section, let us focus on the directed graph corresponding to $\vec{\sigma}_3=(-1,1,-1,1)$, graphically represented in eq.~\eqref{eq:graph_or_5}. We have, identifying $\hat{\mathds{1}}_{\vec{\sigma}_3}$ with the graph $G_{\vec{\sigma_3}}$:
\begin{equation}
\label{eq:or1}
\resizebox{2.cm}{!}{\raisebox{-1.5cm}{
\begin{tikzpicture}

    \begin{feynman}
        \vertex(1);

        \vertex[right = 2cm of 1](2);

        \vertex[below = 2cm of 2](3);

        \vertex[left = 2cm of 3](4);

        \vertex[above right =0.2cm and 0.8cm of 1](L1) {\resizebox{0.5cm}{!}{$e_2$}};
        \vertex[below right =0.2cm and 0.8cm of 4](L2) {\resizebox{0.5cm}{!}{$e_4$}};

        \vertex[below right =0.8cm and 0.2cm of 2](L3) {\resizebox{0.5cm}{!}{$e_3$}};
        \vertex[below left =0.8cm and 0.2cm of 1](L4) {\resizebox{0.5cm}{!}{$e_1$}};

        \vertex[above left =0.05cm and 0.05cm of 1](LV1) {\resizebox{0.5cm}{!}{$v_1$}};
        \vertex[above right =0.05cm and 0.05cm of 2](LV2) {\resizebox{0.5cm}{!}{$v_2$}};
        \vertex[below right =0.05cm and 0.05cm of 3](LV3) {\resizebox{0.5cm}{!}{$v_3$}};
        \vertex[below left =0.05cm and 0.05cm of 4](LV4) {\resizebox{0.5cm}{!}{$v_4$}};

         \diagram*[large]{	
            (1) -- [->-,line width=0.6mm] (2),
            (2) -- [-<-,line width=0.6mm] (3),
            (3) -- [->-,line width=0.6mm] (4),
            (4) -- [-<-,line width=0.6mm] (1),
        }; 
      \path[draw=black, fill=black] (1) circle[radius=0.05];
       \path[draw=black, fill=black] (2) circle[radius=0.05];
       \path[draw=black, fill=black] (3) circle[radius=0.05];
       \path[draw=black, fill=black] (4) circle[radius=0.05];
       
    \end{feynman}
    
\end{tikzpicture}
}}=\hat{\mathds{1}}_{\vec{\sigma}_3}=\int \left[\prod_{j=1}^4\mathrm{d}\tau_j \Theta(\tau_j) e^{i \tilde{E}_j \tau_j}\right] \delta(-\tau_1+\tau_2-\tau_3+\tau_4),
\end{equation}
having redefined $\tilde{E}_j=E_j-(\vec{\sigma}_3)_j p_j^0$. We now relate the process of performing time integrations with the edge-contraction operation. We start by writing
\begin{equation}
\Theta(\tau_1)\Theta(\tau_2)=\Theta(\tau_1-\tau_2)\Theta(\tau_2)+\Theta(\tau_2-\tau_1)\Theta(\tau_1),
\end{equation}
corresponding to a triangulation of the positive orthant $\tau_1,\tau_2>0$ in two cones: $\tau_2>0, \tau_1>\tau_2$ and $\tau_1>0, \tau_2>\tau_1$. Substituted in $\hat{\mathds{1}}_{\vec{\sigma}_3}$, we obtain
\begin{equation}
\resizebox{2.cm}{!}{\raisebox{-1.5cm}{
\begin{tikzpicture}

    \begin{feynman}
        \vertex(1);

        \vertex[right = 2cm of 1](2);

        \vertex[below = 2cm of 2](3);

        \vertex[left = 2cm of 3](4);

        \vertex[above right =0.2cm and 0.8cm of 1](L1) {\resizebox{0.5cm}{!}{$e_2$}};
        \vertex[below right =0.2cm and 0.8cm of 4](L2) {\resizebox{0.5cm}{!}{$e_4$}};

        \vertex[below right =0.8cm and 0.2cm of 2](L3) {\resizebox{0.5cm}{!}{$e_3$}};
        \vertex[below left =0.8cm and 0.2cm of 1](L4) {\resizebox{0.5cm}{!}{$e_1$}};

        \vertex[above left =0.05cm and 0.05cm of 1](LV1) {\resizebox{0.5cm}{!}{$v_1$}};
        \vertex[above right =0.05cm and 0.05cm of 2](LV2) {\resizebox{0.5cm}{!}{$v_2$}};
        \vertex[below right =0.05cm and 0.05cm of 3](LV3) {\resizebox{0.5cm}{!}{$v_3$}};
        \vertex[below left =0.05cm and 0.05cm of 4](LV4) {\resizebox{0.5cm}{!}{$v_4$}};

         \diagram*[large]{	
            (1) -- [->-,line width=0.6mm] (2),
            (2) -- [-<-,line width=0.6mm] (3),
            (3) -- [->-,line width=0.6mm] (4),
            (4) -- [-<-,line width=0.6mm] (1),
        }; 
      \path[draw=black, fill=black] (1) circle[radius=0.05];
       \path[draw=black, fill=black] (2) circle[radius=0.05];
       \path[draw=black, fill=black] (3) circle[radius=0.05];
       \path[draw=black, fill=black] (4) circle[radius=0.05];
       
    \end{feynman}
    
\end{tikzpicture}
}}=\int\left[ \prod_{i=3}^4 \mathrm{d}\tau_i \Theta(\tau_i) e^{i\tilde{E}_i\tau_i}\right] \Bigg[\int  \mathrm{d}\tau_1 \mathrm{d}\tau_2 e^{i\tau_1 \tilde{E}_1+i\tau_2 \tilde{E}_2}\Theta(\tau_1)\Theta(\tau_2-\tau_1)\delta(-\tau_1+\tau_2-\tau_3+\tau_4)\nonumber \end{equation}
\begin{equation}
+\int  \mathrm{d}\tau_1 \mathrm{d}\tau_2 e^{i\tau_1 \tilde{E}_1+i\tau_2 \tilde{E}_2}\Theta(\tau_2)\Theta(\tau_1-\tau_2)\delta(-\tau_1+\tau_2-\tau_3+\tau_4)\Bigg]
\end{equation}
We now change variables $\tau_2'=\tau_2-\tau_1$, $\tau_j'=\tau_j$, $j=1,3,4$ for the first integral and $\tau_1'=\tau_1-\tau_2$, $\tau_j'=\tau_j$, $j=2,3,4$ for the second one
\begin{equation}
\resizebox{2.cm}{!}{\raisebox{-1.5cm}{
\begin{tikzpicture}

    \begin{feynman}
        \vertex(1);

        \vertex[right = 2cm of 1](2);

        \vertex[below = 2cm of 2](3);

        \vertex[left = 2cm of 3](4);

        \vertex[above right =0.2cm and 0.8cm of 1](L1) {\resizebox{0.5cm}{!}{$e_2$}};
        \vertex[below right =0.2cm and 0.8cm of 4](L2) {\resizebox{0.5cm}{!}{$e_4$}};

        \vertex[below right =0.8cm and 0.2cm of 2](L3) {\resizebox{0.5cm}{!}{$e_3$}};
        \vertex[below left =0.8cm and 0.2cm of 1](L4) {\resizebox{0.5cm}{!}{$e_1$}};

        \vertex[above left =0.05cm and 0.05cm of 1](LV1) {\resizebox{0.5cm}{!}{$v_1$}};
        \vertex[above right =0.05cm and 0.05cm of 2](LV2) {\resizebox{0.5cm}{!}{$v_2$}};
        \vertex[below right =0.05cm and 0.05cm of 3](LV3) {\resizebox{0.5cm}{!}{$v_3$}};
        \vertex[below left =0.05cm and 0.05cm of 4](LV4) {\resizebox{0.5cm}{!}{$v_4$}};

         \diagram*[large]{	
            (1) -- [->-,line width=0.6mm] (2),
            (2) -- [-<-,line width=0.6mm] (3),
            (3) -- [->-,line width=0.6mm] (4),
            (4) -- [-<-,line width=0.6mm] (1),
        }; 
      \path[draw=black, fill=black] (1) circle[radius=0.05];
       \path[draw=black, fill=black] (2) circle[radius=0.05];
       \path[draw=black, fill=black] (3) circle[radius=0.05];
       \path[draw=black, fill=black] (4) circle[radius=0.05];
       
    \end{feynman}
    
\end{tikzpicture}
}}=\int\left[ \prod_{i\in\{2,3,4\}} \mathrm{d}\tau_i \Theta(\tau_i) e^{i\tilde{E}_i\tau_i}\right]\delta(\tau_2-\tau_3+\tau_4) \int  \mathrm{d}\tau_1 \Theta(\tau_1) e^{i\tau_1 (\tilde{E}_1+\tilde{E}_2)} \nonumber \end{equation}
\begin{equation}
+\int \left[\prod_{i\in\{1,3,4\}} \mathrm{d}\tau_i \Theta(\tau_i) e^{i\tilde{E}_i\tau_i}\right]\delta(-\tau_1-\tau_3+\tau_4) \int  \mathrm{d}\tau_2 \Theta(\tau_2) e^{i\tau_2 (\tilde{E}_1+\tilde{E}_2)}.
\end{equation}
The nested integrations are now trivial to perform
\begin{equation}
\resizebox{2.cm}{!}{\raisebox{-1.5cm}{
\begin{tikzpicture}

    \begin{feynman}
        \vertex(1);

        \vertex[right = 2cm of 1](2);

        \vertex[below = 2cm of 2](3);

        \vertex[left = 2cm of 3](4);

        \vertex[above right =0.2cm and 0.8cm of 1](L1) {\resizebox{0.5cm}{!}{$e_2$}};
        \vertex[below right =0.2cm and 0.8cm of 4](L2) {\resizebox{0.5cm}{!}{$e_4$}};

        \vertex[below right =0.8cm and 0.2cm of 2](L3) {\resizebox{0.5cm}{!}{$e_3$}};
        \vertex[below left =0.8cm and 0.2cm of 1](L4) {\resizebox{0.5cm}{!}{$e_1$}};

        \vertex[above left =0.05cm and 0.05cm of 1](LV1) {\resizebox{0.5cm}{!}{$v_1$}};
        \vertex[above right =0.05cm and 0.05cm of 2](LV2) {\resizebox{0.5cm}{!}{$v_2$}};
        \vertex[below right =0.05cm and 0.05cm of 3](LV3) {\resizebox{0.5cm}{!}{$v_3$}};
        \vertex[below left =0.05cm and 0.05cm of 4](LV4) {\resizebox{0.5cm}{!}{$v_4$}};

         \diagram*[large]{	
            (1) -- [->-,line width=0.6mm] (2),
            (2) -- [-<-,line width=0.6mm] (3),
            (3) -- [->-,line width=0.6mm] (4),
            (4) -- [-<-,line width=0.6mm] (1),
        }; 
      \path[draw=black, fill=black] (1) circle[radius=0.05];
       \path[draw=black, fill=black] (2) circle[radius=0.05];
       \path[draw=black, fill=black] (3) circle[radius=0.05];
       \path[draw=black, fill=black] (4) circle[radius=0.05];
       
    \end{feynman}
    
\end{tikzpicture}
}}=\frac{i}{\tilde{E}_1+\tilde{E}_2}\Bigg[\int\left[ \prod_{i\in\{2,3,4\}} \mathrm{d}\tau_i \Theta(\tau_i) e^{i\tilde{E}_i\tau_i}\right]\delta(\tau_2-\tau_3+\tau_4) \nonumber  \end{equation}
\begin{equation}
+\int \left[\prod_{i\in\{1,3,4\}} \mathrm{d}\tau_i \Theta(\tau_i) e^{i\tilde{E}_i\tau_i}\right]\delta(-\tau_1-\tau_3+\tau_4) \Bigg].
\end{equation}
Such an equation has a straight-forward diagrammatic representation. In particular, we recognise that it can be written as
\begin{equation}
\resizebox{2.cm}{!}{\raisebox{-1.5cm}{
\begin{tikzpicture}

    \begin{feynman}
        \vertex(1);

        \vertex[right = 2cm of 1](2);

        \vertex[below = 2cm of 2](3);

        \vertex[left = 2cm of 3](4);

        \vertex[above right =0.2cm and 0.8cm of 1](L1) {\resizebox{0.5cm}{!}{$e_2$}};
        \vertex[below right =0.2cm and 0.8cm of 4](L2) {\resizebox{0.5cm}{!}{$e_4$}};

        \vertex[below right =0.8cm and 0.2cm of 2](L3) {\resizebox{0.5cm}{!}{$e_3$}};
        \vertex[below left =0.8cm and 0.2cm of 1](L4) {\resizebox{0.5cm}{!}{$e_1$}};

        \vertex[above left =0.05cm and 0.05cm of 1](LV1) {\resizebox{0.5cm}{!}{$v_1$}};
        \vertex[above right =0.05cm and 0.05cm of 2](LV2) {\resizebox{0.5cm}{!}{$v_2$}};
        \vertex[below right =0.05cm and 0.05cm of 3](LV3) {\resizebox{0.5cm}{!}{$v_3$}};
        \vertex[below left =0.05cm and 0.05cm of 4](LV4) {\resizebox{0.5cm}{!}{$v_4$}};

         \diagram*[large]{	
            (1) -- [->-,line width=0.6mm] (2),
            (2) -- [-<-,line width=0.6mm] (3),
            (3) -- [->-,line width=0.6mm] (4),
            (4) -- [-<-,line width=0.6mm] (1),
        }; 
      \path[draw=black, fill=black] (1) circle[radius=0.05];
       \path[draw=black, fill=black] (2) circle[radius=0.05];
       \path[draw=black, fill=black] (3) circle[radius=0.05];
       \path[draw=black, fill=black] (4) circle[radius=0.05];
       
    \end{feynman}
    
\end{tikzpicture}
}}=\frac{i}{(\tilde{E}_1+\tilde{E}_2)}\left[\raisebox{-0.6cm}{\resizebox{1.6cm}{!}{
\begin{tikzpicture}

    \begin{feynman}
        \vertex(1);

        \vertex[right = 2cm of 1](2);

        \vertex[below = 2cm of 2](3);

        \vertex[left = 2cm of 3](4);

        \vertex[above right =0.8cm and 0.2cm of 3](L1) {\resizebox{0.5cm}{!}{$e_3$}};
        \vertex[below left =0.2cm and 0.8cm of 3](L2) {\resizebox{0.5cm}{!}{$e_4$}};

        \vertex[above right =1cm and 0.1cm of 4](L3) {\resizebox{0.5cm}{!}{$e_1$}};

        \vertex[above right =0.05cm and 0.05cm of 2](LV2) {\resizebox{0.7cm}{!}{$v_{12}$}};
        \vertex[below right =0.05cm and 0.05cm of 3](LV3) {\resizebox{0.5cm}{!}{$v_3$}};
        \vertex[below left =0.05cm and 0.05cm of 4](LV4) {\resizebox{0.5cm}{!}{$v_4$}};

         \diagram*[large]{	
            (2) -- [-<-,line width=0.6mm] (3),
            (3) -- [->-,line width=0.6mm] (4),
            (4) -- [-<-,line width=0.6mm] (2),
        }; 
       \path[draw=black, fill=black] (2) circle[radius=0.05];
       \path[draw=black, fill=black] (3) circle[radius=0.05];
       \path[draw=black, fill=black] (4) circle[radius=0.05];
       
    \end{feynman}
    
\end{tikzpicture}
}}+\raisebox{-0.6cm}{\resizebox{1.6cm}{!}{
\begin{tikzpicture}

    \begin{feynman}
        \vertex(1);

        \vertex[right = 2cm of 1](2);

        \vertex[below = 2cm of 2](3);

        \vertex[left = 2cm of 3](4);

        \vertex[above right =0.8cm and 0.2cm of 3](L1) {\resizebox{0.5cm}{!}{$e_3$}};
        \vertex[below left =0.2cm and 0.8cm of 3](L2) {\resizebox{0.5cm}{!}{$e_4$}};

        \vertex[above right =1cm and 0.1cm of 4](L3) {\resizebox{0.5cm}{!}{$e_2$}};

        \vertex[above right =0.05cm and 0.05cm of 2](LV2) {\resizebox{0.5cm}{!}{$v_2$}};
        \vertex[below right =0.05cm and 0.05cm of 3](LV3) {\resizebox{0.5cm}{!}{$v_{3}$}};
        \vertex[below left =0.05cm and 0.05cm of 4](LV4) {\resizebox{0.7cm}{!}{$v_{14}$}};

         \diagram*[large]{	
            (2) -- [-<-,line width=0.6mm] (3),
            (3) -- [->-,line width=0.6mm] (4),
            (4) -- [->-,line width=0.6mm] (2),
        }; 
       \path[draw=black, fill=black] (2) circle[radius=0.05];
       \path[draw=black, fill=black] (3) circle[radius=0.05];
       \path[draw=black, fill=black] (4) circle[radius=0.05];
       
    \end{feynman}
    
\end{tikzpicture}
}}\right].
\end{equation}
We thus recognise that the overall effect of having performed the integration in the way we have laid down is, at the graph level, equivalent to contracting one-by-one the two edges $e_1$ and $e_2$ adjacent to the vertex $v_1$. It turns out that the operation can be iterated. We can now choose one vertex for each of the two contracted graphs: for the first, we choose the vertex $v_4$ common to $e_1$ and $e_4$; for the second, we choose the vertex $v_2$ common to $e_2$ and $e_3$. Perfoming the contraction, we obtain:
\begin{equation}
\resizebox{2.cm}{!}{\raisebox{-1.5cm}{
\begin{tikzpicture}

    \begin{feynman}
        \vertex(1);

        \vertex[right = 2cm of 1](2);

        \vertex[below = 2cm of 2](3);

        \vertex[left = 2cm of 3](4);

        \vertex[above right =0.2cm and 0.8cm of 1](L1) {\resizebox{0.5cm}{!}{$e_2$}};
        \vertex[below right =0.2cm and 0.8cm of 4](L2) {\resizebox{0.5cm}{!}{$e_4$}};

        \vertex[below right =0.8cm and 0.2cm of 2](L3) {\resizebox{0.5cm}{!}{$e_3$}};
        \vertex[below left =0.8cm and 0.2cm of 1](L4) {\resizebox{0.5cm}{!}{$e_1$}};

        \vertex[above left =0.05cm and 0.05cm of 1](LV1) {\resizebox{0.5cm}{!}{$v_1$}};
        \vertex[above right =0.05cm and 0.05cm of 2](LV2) {\resizebox{0.5cm}{!}{$v_2$}};
        \vertex[below right =0.05cm and 0.05cm of 3](LV3) {\resizebox{0.5cm}{!}{$v_3$}};
        \vertex[below left =0.05cm and 0.05cm of 4](LV4) {\resizebox{0.5cm}{!}{$v_4$}};

         \diagram*[large]{	
            (1) -- [->-,line width=0.6mm] (2),
            (2) -- [-<-,line width=0.6mm] (3),
            (3) -- [->-,line width=0.6mm] (4),
            (4) -- [-<-,line width=0.6mm] (1),
        }; 
      \path[draw=black, fill=black] (1) circle[radius=0.05];
       \path[draw=black, fill=black] (2) circle[radius=0.05];
       \path[draw=black, fill=black] (3) circle[radius=0.05];
       \path[draw=black, fill=black] (4) circle[radius=0.05];
       
    \end{feynman}
    
\end{tikzpicture}
}}=\frac{i}{(\tilde{E}_1+\tilde{E}_2)}\Bigg[\frac{i}{(\tilde{E}_1+\tilde{E}_4)}\Bigg[\resizebox{1cm}{!}{\raisebox{-1.2cm}{
\begin{tikzpicture}

    \begin{feynman}
        \vertex(1);
        \vertex[below = 2cm of 1](3);

        \vertex[below left = 1cm and 0.3cm of 1](L1) {$e_3$};
        \vertex[below right = 1cm and 0.3cm of 1](L2) {$e_4$};

        \vertex[above =0.1cm of 1](LV1) {\resizebox{0.7cm}{!}{$v_{124}$}};
        \vertex[below =0.1cm of 3](LV2) {$v_3$};

         \diagram*[large]{	
            (1) -- [-<-,line width=0.6mm, out=-60, in=60] (3),
            (1) -- [-<-,line width=0.6mm, out=240, in=120] (3),
        }; 
       \path[draw=black, fill=black] (1) circle[radius=0.05];
       \path[draw=black, fill=black] (3) circle[radius=0.05];
       
    \end{feynman}
    
\end{tikzpicture}
}}+\resizebox{1cm}{!}{\raisebox{-1.2cm}{
\begin{tikzpicture}

    \begin{feynman}
        \vertex(1);
        \vertex[below = 2cm of 1](3);

        \vertex[below left = 1cm and 0.3cm of 1](L1) {$e_1$};
        \vertex[below right = 1cm and 0.3cm of 1](L1) {$e_3$};

        \vertex[above =0.1cm of 1](LV1) {\resizebox{0.55cm}{!}{$v_{12}$}};
        \vertex[below =0.1cm of 3](LV2) {\resizebox{0.55cm}{!}{$v_{34}$}};

         \diagram*[large]{	
            (1) -- [-<-,line width=0.6mm, out=-60, in=60] (3),
            (1) -- [->-,line width=0.6mm, out=240, in=120] (3),
        }; 
       \path[draw=black, fill=black] (1) circle[radius=0.05];
       \path[draw=black, fill=black] (3) circle[radius=0.05];
       
    \end{feynman}
    
\end{tikzpicture}
}}\Bigg] \end{equation}
\begin{equation}
+\frac{i}{(\tilde{E}_2+\tilde{E}_3)}\Bigg[\resizebox{1cm}{!}{\raisebox{-1.2cm}{
\begin{tikzpicture}

    \begin{feynman}
        \vertex(1);
        \vertex[below = 2cm of 1](3);

        \vertex[below left = 1cm and 0.3cm of 1](L1) {$e_3$};
        \vertex[below right = 1cm and 0.3cm of 1](L1) {$e_4$};

        \vertex[above =0.1cm of 1](LV1) {\resizebox{0.7cm}{!}{$v_{124}$}};
        \vertex[below =0.1cm of 3](LV2) {$v_3$};

         \diagram*[large]{	
            (1) -- [-<-,line width=0.6mm, out=-60, in=60] (3),
            (1) -- [-<-,line width=0.6mm, out=240, in=120] (3),
        }; 
       \path[draw=black, fill=black] (1) circle[radius=0.05];
       \path[draw=black, fill=black] (3) circle[radius=0.05];
       
    \end{feynman}
    
\end{tikzpicture}
}}+\resizebox{1cm}{!}{\raisebox{-1.2cm}{
\begin{tikzpicture}

    \begin{feynman}
        \vertex(1);
        \vertex[below = 2cm of 1](3);

        \vertex[below left = 1cm and 0.3cm of 1](L1) {$e_2$};
        \vertex[below right = 1cm and 0.3cm of 1](L1) {$e_4$};

        \vertex[above =0.1cm of 1](LV1) {\resizebox{0.55cm}{!}{$v_{14}$}};
        \vertex[below =0.1cm of 3](LV2) {\resizebox{0.55cm}{!}{$v_{23}$}};

         \diagram*[large]{	
            (1) -- [-<-,line width=0.6mm, out=-60, in=60] (3),
            (1) -- [->-,line width=0.6mm, out=240, in=120] (3),
        }; 
       \path[draw=black, fill=black] (1) circle[radius=0.05];
       \path[draw=black, fill=black] (3) circle[radius=0.05];
       
    \end{feynman}
    
\end{tikzpicture}
}}\Bigg]\Bigg].
\end{equation}
We now notice that 
\begin{equation}
\raisebox{-0.4cm}{\resizebox{1.7cm}{!}{\includegraphics[page=16]{tikz.pdf}}}\hspace{-0.5cm}=\hspace{-0.2cm}\raisebox{-0.4cm}{\resizebox{1.7cm}{!}{\includegraphics[page=19]{tikz.pdf}}}\hspace{-0.5cm}=0,
\end{equation}
since they are oriented in such a way to allow for a directed cycle. In other words, since these two graphs are not acyclic, the corresponding integrals vanish, because the domain of integration is empty. This implies that 
\begin{equation}
\resizebox{2.cm}{!}{\raisebox{-1.5cm}{
\begin{tikzpicture}

    \begin{feynman}
        \vertex(1);

        \vertex[right = 2cm of 1](2);

        \vertex[below = 2cm of 2](3);

        \vertex[left = 2cm of 3](4);

        \vertex[above right =0.2cm and 0.8cm of 1](L1) {\resizebox{0.5cm}{!}{$e_2$}};
        \vertex[below right =0.2cm and 0.8cm of 4](L2) {\resizebox{0.5cm}{!}{$e_4$}};

        \vertex[below right =0.8cm and 0.2cm of 2](L3) {\resizebox{0.5cm}{!}{$e_3$}};
        \vertex[below left =0.8cm and 0.2cm of 1](L4) {\resizebox{0.5cm}{!}{$e_1$}};

        \vertex[above left =0.05cm and 0.05cm of 1](LV1) {\resizebox{0.5cm}{!}{$v_1$}};
        \vertex[above right =0.05cm and 0.05cm of 2](LV2) {\resizebox{0.5cm}{!}{$v_2$}};
        \vertex[below right =0.05cm and 0.05cm of 3](LV3) {\resizebox{0.5cm}{!}{$v_3$}};
        \vertex[below left =0.05cm and 0.05cm of 4](LV4) {\resizebox{0.5cm}{!}{$v_4$}};

         \diagram*[large]{	
            (1) -- [->-,line width=0.6mm] (2),
            (2) -- [-<-,line width=0.6mm] (3),
            (3) -- [->-,line width=0.6mm] (4),
            (4) -- [-<-,line width=0.6mm] (1),
        }; 
      \path[draw=black, fill=black] (1) circle[radius=0.05];
       \path[draw=black, fill=black] (2) circle[radius=0.05];
       \path[draw=black, fill=black] (3) circle[radius=0.05];
       \path[draw=black, fill=black] (4) circle[radius=0.05];
       
    \end{feynman}
    
\end{tikzpicture}
}}=\frac{i}{(\tilde{E}_1+\tilde{E}_2)}\left[\frac{i}{(\tilde{E}_1+\tilde{E}_4)}\resizebox{1cm}{!}{\raisebox{-1.2cm}{
\begin{tikzpicture}

    \begin{feynman}
        \vertex(1);
        \vertex[below = 2cm of 1](3);

        \vertex[below left = 1cm and 0.3cm of 1](L1) {$e_3$};
        \vertex[below right = 1cm and 0.3cm of 1](L2) {$e_4$};

        \vertex[above =0.1cm of 1](LV1) {\resizebox{0.7cm}{!}{$v_{124}$}};
        \vertex[below =0.1cm of 3](LV2) {$v_3$};

         \diagram*[large]{	
            (1) -- [-<-,line width=0.6mm, out=-60, in=60] (3),
            (1) -- [-<-,line width=0.6mm, out=240, in=120] (3),
        }; 
       \path[draw=black, fill=black] (1) circle[radius=0.05];
       \path[draw=black, fill=black] (3) circle[radius=0.05];
       
    \end{feynman}
    
\end{tikzpicture}
}}+\frac{i}{(\tilde{E}_2+\tilde{E}_3)}\resizebox{1cm}{!}{\raisebox{-1.2cm}{
\begin{tikzpicture}

    \begin{feynman}
        \vertex(1);
        \vertex[below = 2cm of 1](3);

        \vertex[below left = 1cm and 0.3cm of 1](L1) {$e_3$};
        \vertex[below right = 1cm and 0.3cm of 1](L1) {$e_4$};

        \vertex[above =0.1cm of 1](LV1) {\resizebox{0.7cm}{!}{$v_{124}$}};
        \vertex[below =0.1cm of 3](LV2) {$v_3$};

         \diagram*[large]{	
            (1) -- [-<-,line width=0.6mm, out=-60, in=60] (3),
            (1) -- [-<-,line width=0.6mm, out=240, in=120] (3),
        }; 
       \path[draw=black, fill=black] (1) circle[radius=0.05];
       \path[draw=black, fill=black] (3) circle[radius=0.05];
       
    \end{feynman}
    
\end{tikzpicture}
}}\right].
\end{equation}
Finally, we compute the two bubble integrals explicitly. We have 
\begin{equation}
\resizebox{1cm}{!}{\raisebox{-1.2cm}{
\begin{tikzpicture}

    \begin{feynman}
        \vertex(1);
        \vertex[below = 2cm of 1](3);

        \vertex[below left = 1cm and 0.3cm of 1](L1) {$e_3$};
        \vertex[below right = 1cm and 0.3cm of 1](L1) {$e_4$};

        \vertex[above =0.1cm of 1](LV1) {\resizebox{0.7cm}{!}{$v_{124}$}};
        \vertex[below =0.1cm of 3](LV2) {$v_3$};

         \diagram*[large]{	
            (1) -- [-<-,line width=0.6mm, out=-60, in=60] (3),
            (1) -- [-<-,line width=0.6mm, out=240, in=120] (3),
        }; 
       \path[draw=black, fill=black] (1) circle[radius=0.05];
       \path[draw=black, fill=black] (3) circle[radius=0.05];
       
    \end{feynman}
    
\end{tikzpicture}
}}=\int \mathrm{d}\tau_3 \mathrm{d}\tau_4 \Theta(\tau_3)\Theta(\tau_4) \delta(\tau_3-\tau_4) e^{i\tau_3 \tilde{E}_3+i\tau_4 \tilde{E}_4}=\frac{i}{\tilde{E}_3+\tilde{E}_4}. \label{eq:last_step}
\end{equation}
Finally, we can write:
\begin{equation}
\label{eq:end_contraction}
\resizebox{2.cm}{!}{\raisebox{-1.5cm}{
\begin{tikzpicture}

    \begin{feynman}
        \vertex(1);

        \vertex[right = 2cm of 1](2);

        \vertex[below = 2cm of 2](3);

        \vertex[left = 2cm of 3](4);

        \vertex[above right =0.2cm and 0.8cm of 1](L1) {\resizebox{0.5cm}{!}{$e_2$}};
        \vertex[below right =0.2cm and 0.8cm of 4](L2) {\resizebox{0.5cm}{!}{$e_4$}};

        \vertex[below right =0.8cm and 0.2cm of 2](L3) {\resizebox{0.5cm}{!}{$e_3$}};
        \vertex[below left =0.8cm and 0.2cm of 1](L4) {\resizebox{0.5cm}{!}{$e_1$}};

        \vertex[above left =0.05cm and 0.05cm of 1](LV1) {\resizebox{0.5cm}{!}{$v_1$}};
        \vertex[above right =0.05cm and 0.05cm of 2](LV2) {\resizebox{0.5cm}{!}{$v_2$}};
        \vertex[below right =0.05cm and 0.05cm of 3](LV3) {\resizebox{0.5cm}{!}{$v_3$}};
        \vertex[below left =0.05cm and 0.05cm of 4](LV4) {\resizebox{0.5cm}{!}{$v_4$}};

         \diagram*[large]{	
            (1) -- [->-,line width=0.6mm] (2),
            (2) -- [-<-,line width=0.6mm] (3),
            (3) -- [->-,line width=0.6mm] (4),
            (4) -- [-<-,line width=0.6mm] (1),
        }; 
      \path[draw=black, fill=black] (1) circle[radius=0.05];
       \path[draw=black, fill=black] (2) circle[radius=0.05];
       \path[draw=black, fill=black] (3) circle[radius=0.05];
       \path[draw=black, fill=black] (4) circle[radius=0.05];
       
    \end{feynman}
    
\end{tikzpicture}
}}=\frac{i}{(\tilde{E}_1+\tilde{E}_2)}\left[\frac{i^2}{(\tilde{E}_1+\tilde{E}_4)(\tilde{E}_3+\tilde{E}_4)}+\frac{i^2}{(\tilde{E}_2+\tilde{E}_3)(\tilde{E}_3+\tilde{E}_4)}\right].
\end{equation}
This concludes the computation of the Fourier transform $\hat{\mathds{1}}_{\vec{\sigma}_3}$. The edge-contraction operation can be used to compute $\hat{\mathds{1}}_{\vec{\sigma}}$ for any orientation $\vec{\sigma}$. 

\subsection{Aside on parallel edges}

\noindent We can also see eq.~\eqref{eq:last_step}, in which we performed the integration of the bubble explicitly, as a result of the contraction operation with vertex choice given by $v_3$: whenever two or more edges connect the same two vertices, they must be contracted all at once (see ref.~\cite{Capatti:2022mly} for details). In a formula:
\begin{equation}
\label{eq:end_contraction}
\resizebox{1.5cm}{!}{\raisebox{-1.5cm}{
\begin{tikzpicture}

    \begin{feynman}
        \vertex(1);

        \vertex[below = 2cm of 1](2);

        \vertex[above =0.1cm of 1](L1) {\resizebox{0.5cm}{!}{$v_1$}};
        \vertex[below =0.1cm of 2](L2) {\resizebox{0.5cm}{!}{$v_2$}};

        \vertex[above left = 0.4cm and 0.75cm of 2](L2) {\resizebox{0.5cm}{!}{$e_1$}};

        \vertex[above left = 0.4cm and 0.12cm of 2](L2) {\resizebox{0.5cm}{!}{$e_2$}};

        \vertex[above right = 0.4cm and 0.8cm of 2](L2) {\resizebox{0.5cm}{!}{$e_n$}};

        \vertex[above right = 0.5cm and 0.cm of 2](L2) {\resizebox{0.4cm}{!}{$\boldsymbol{\ldots}$}};

         \diagram*[large]{	
            (1) -- [->-,line width=0.6mm, half right] (2),
            (1) -- [->-,line width=0.6mm, out=-115, in=115] (2),
            (1) -- [->-,line width=0.6mm, half left] (2),

        }; 
      \path[draw=black, fill=black] (1) circle[radius=0.05];
       \path[draw=black, fill=black] (2) circle[radius=0.05];

    \end{feynman}
    
\end{tikzpicture}
}}=\frac{i}{\sum_{i=1}^n \tilde{E}_i} \resizebox{0.6cm}{!}{\raisebox{0.05cm}{
\begin{tikzpicture}

    \begin{feynman}
        \vertex(1);

        \vertex[above =0.1cm of 1](L1) {\resizebox{0.75cm}{!}{$v_{12}$}};
        
      \path[draw=black, fill=black] (1) circle[radius=0.05];

    \end{feynman}
    
\end{tikzpicture}
}}=\frac{i}{\sum_{i=1}^n \tilde{E}_i}.
\end{equation}
This relation is used independently of whether it appears at the last iteration of the edge-contraction procedure or not. The general rule, for an arbitrarily complicated graph, is the following: if, at any point, the choice of vertex requires to contract parallel edges, they should be contracted all at once. Consider, for example, the following orientation of the dunce's hat
\begin{equation}
\label{eq:end_contraction}
\resizebox{1.5cm}{!}{\raisebox{-1.5cm}{
\begin{tikzpicture}

    \begin{feynman}
        \vertex(1);

        \vertex[right = 2cm of 1](2);

        \vertex[below right = 2cm and 1cm of 1](3);

        \vertex[above =0.1cm of 1](L1) {\resizebox{0.5cm}{!}{$v_1$}};
        \vertex[above =0.1cm of 2](L2) {\resizebox{0.5cm}{!}{$v_2$}};

        \vertex[below =0.1cm of 3](L3) {\resizebox{0.5cm}{!}{$v_3$}};

        \vertex[above left = 0.5cm and 0.6cm of 2](L2) {\resizebox{0.5cm}{!}{$e_1$}};

        \vertex[below left = 0.5cm and 0.6cm of 2](L2) {\resizebox{0.5cm}{!}{$e_2$}};

        \vertex[below left = 0.8cm and -0.35cm of 2](L2) {\resizebox{0.5cm}{!}{$e_4$}};

        \vertex[below right = 0.8cm and -0.3cm of 1](L2) {\resizebox{0.5cm}{!}{$e_3$}};

         \diagram*[large]{	
            (1) -- [->-,line width=0.6mm, out=45, in=135] (2),
            
            (1) -- [->-,line width=0.6mm, out=-45, in=-135] (2),

            (1) -- [->-,line width=0.6mm] (3),

            (2) -- [->-,line width=0.6mm] (3),

        }; 
      \path[draw=black, fill=black] (1) circle[radius=0.05];
       \path[draw=black, fill=black] (2) circle[radius=0.05];
       \path[draw=black, fill=black] (3) circle[radius=0.05];

    \end{feynman}
    
\end{tikzpicture}
}}=\frac{i}{\tilde{E}_1+\tilde{E}_2+\tilde{E}_3}\Bigg[\resizebox{1.cm}{!}{\raisebox{-1.5cm}{\begin{tikzpicture}

    \begin{feynman}
        \vertex(1);
        \vertex[below = 2cm of 1](3);

        \vertex[below left = 1cm and 0.3cm of 1](L1) {$e_3$};
        \vertex[below right = 1cm and 0.3cm of 1](L1) {$e_4$};

        \vertex[above =0.1cm of 1](LV1) {\resizebox{0.7cm}{!}{$v_{12}$}};
        \vertex[below =0.1cm of 3](LV2) {$v_3$};

         \diagram*[large]{	
            (1) -- [->-,line width=0.6mm, out=-60, in=60] (3),
            (1) -- [->-,line width=0.6mm, out=240, in=120] (3),
        }; 
       \path[draw=black, fill=black] (1) circle[radius=0.05];
       \path[draw=black, fill=black] (3) circle[radius=0.05];
       
    \end{feynman}
    
\end{tikzpicture}}}+\resizebox{1.5cm}{!}{\raisebox{-1.5cm}{\begin{tikzpicture}

    \begin{feynman}
        \vertex(1);
        \vertex[below = 2cm of 1](3);

        \vertex[below left = 1cm and 0.3cm of 1](L1) {$e_1$};
        \vertex[below right = 1cm and 0.8cm of 1](L1) {$e_2$};
        \vertex[below right = 1cm and 0.cm of 1](L1) {$e_4$};

        \vertex[above =0.1cm of 1](LV1) {\resizebox{0.7cm}{!}{$v_{13}$}};
        \vertex[below =0.1cm of 3](LV2) {$v_2$};

         \diagram*[large]{	
            (1) -- [->-,line width=0.6mm, half right] (3),
            (1) -- [->-,line width=0.6mm, half left] (3),
            (1) -- [-<-,line width=0.6mm] (3),
        }; 
       \path[draw=black, fill=black] (1) circle[radius=0.05];
       \path[draw=black, fill=black] (3) circle[radius=0.05];
       
    \end{feynman}
    
\end{tikzpicture}}}\Bigg]
\end{equation}
having chosen the vertex $v_1$ to perform edge-contraction. The second edge-contracted graph has a directed cycle. Thus, it evaluates to zero, and we can iterate the edge-contraction operation
\begin{equation}
\label{eq:end_contraction}
\resizebox{1.5cm}{!}{\raisebox{-1.5cm}{
\begin{tikzpicture}

    \begin{feynman}
        \vertex(1);

        \vertex[right = 2cm of 1](2);

        \vertex[below right = 2cm and 1cm of 1](3);

        \vertex[above =0.1cm of 1](L1) {\resizebox{0.5cm}{!}{$v_1$}};
        \vertex[above =0.1cm of 2](L2) {\resizebox{0.5cm}{!}{$v_2$}};

        \vertex[below =0.1cm of 3](L3) {\resizebox{0.5cm}{!}{$v_3$}};

        \vertex[above left = 0.5cm and 0.6cm of 2](L2) {\resizebox{0.5cm}{!}{$e_1$}};

        \vertex[below left = 0.5cm and 0.6cm of 2](L2) {\resizebox{0.5cm}{!}{$e_2$}};

        \vertex[below left = 0.8cm and -0.35cm of 2](L2) {\resizebox{0.5cm}{!}{$e_4$}};

        \vertex[below right = 0.8cm and -0.3cm of 1](L2) {\resizebox{0.5cm}{!}{$e_3$}};

         \diagram*[large]{	
            (1) -- [->-,line width=0.6mm, out=45, in=135] (2),
            
            (1) -- [->-,line width=0.6mm, out=-45, in=-135] (2),

            (1) -- [->-,line width=0.6mm] (3),

            (2) -- [->-,line width=0.6mm] (3),

        }; 
      \path[draw=black, fill=black] (1) circle[radius=0.05];
       \path[draw=black, fill=black] (2) circle[radius=0.05];
       \path[draw=black, fill=black] (3) circle[radius=0.05];

    \end{feynman}
    
\end{tikzpicture}
}}=\frac{i}{\tilde{E}_1+\tilde{E}_2+\tilde{E}_3}\resizebox{1.cm}{!}{\raisebox{-1.5cm}{\begin{tikzpicture}

    \begin{feynman}
        \vertex(1);
        \vertex[below = 2cm of 1](3);

        \vertex[below left = 1cm and 0.3cm of 1](L1) {$e_3$};
        \vertex[below right = 1cm and 0.3cm of 1](L1) {$e_4$};

        \vertex[above =0.1cm of 1](LV1) {\resizebox{0.7cm}{!}{$v_{12}$}};
        \vertex[below =0.1cm of 3](LV2) {$v_3$};

         \diagram*[large]{	
            (1) -- [->-,line width=0.6mm, out=-60, in=60] (3),
            (1) -- [->-,line width=0.6mm, out=240, in=120] (3),
        }; 
       \path[draw=black, fill=black] (1) circle[radius=0.05];
       \path[draw=black, fill=black] (3) circle[radius=0.05];
       
    \end{feynman}
    
\end{tikzpicture}}}=\frac{i}{\tilde{E}_1+\tilde{E}_2+\tilde{E}_3}\frac{i}{\tilde{E}_3+\tilde{E}_4}.
\end{equation}
Having understood the treatment of parallel edges, we return to the box example.

\subsection{Another example}

\noindent We now look at another orientation, corresponding to $\vec{\sigma}_4=(-1,1,-1,-1)$. Again, we may define $\tilde{E}_i=E_i-(\vec{\sigma}_4)_i p_i^0$, and write
\begin{equation}
\label{eq:or2}
\resizebox{2.cm}{!}{\raisebox{-1.5cm}{
\begin{tikzpicture}

    \begin{feynman}
        \vertex(1);

        \vertex[right = 2cm of 1](2);

        \vertex[below = 2cm of 2](3);

        \vertex[left = 2cm of 3](4);

        \vertex[above right =0.2cm and 0.8cm of 1](L1) {\resizebox{0.5cm}{!}{$e_2$}};
        \vertex[below right =0.2cm and 0.8cm of 4](L2) {\resizebox{0.5cm}{!}{$e_4$}};

        \vertex[below right =0.8cm and 0.2cm of 2](L3) {\resizebox{0.5cm}{!}{$e_3$}};
        \vertex[below left =0.8cm and 0.2cm of 1](L4) {\resizebox{0.5cm}{!}{$e_1$}};

        \vertex[above left =0.05cm and 0.05cm of 1](LV1) {\resizebox{0.5cm}{!}{$v_1$}};
        \vertex[above right =0.05cm and 0.05cm of 2](LV2) {\resizebox{0.5cm}{!}{$v_2$}};
        \vertex[below right =0.05cm and 0.05cm of 3](LV3) {\resizebox{0.5cm}{!}{$v_3$}};
        \vertex[below left =0.05cm and 0.05cm of 4](LV4) {\resizebox{0.5cm}{!}{$v_4$}};

         \diagram*[large]{	
            (1) -- [->-,line width=0.6mm] (2),
            (2) -- [-<-,line width=0.6mm] (3),
            (3) -- [-<-,line width=0.6mm] (4),
            (4) -- [-<-,line width=0.6mm] (1),
        }; 
      \path[draw=black, fill=black] (1) circle[radius=0.05];
       \path[draw=black, fill=black] (2) circle[radius=0.05];
       \path[draw=black, fill=black] (3) circle[radius=0.05];
       \path[draw=black, fill=black] (4) circle[radius=0.05];
       
    \end{feynman}
    
\end{tikzpicture}
}}=\int \left[\prod_{j=1}^4\mathrm{d}\tau_j \Theta(\tau_j) e^{i \tilde{E}_j \tau_j}\right] \delta(-\tau_1+\tau_2-\tau_3-\tau_4).
\end{equation}
Let us apply the contraction operation. We start from $v_1$ and obtain
\begin{equation}
\resizebox{2.cm}{!}{\raisebox{-1.5cm}{
\begin{tikzpicture}

    \begin{feynman}
        \vertex(1);

        \vertex[right = 2cm of 1](2);

        \vertex[below = 2cm of 2](3);

        \vertex[left = 2cm of 3](4);

        \vertex[above right =0.2cm and 0.8cm of 1](L1) {\resizebox{0.5cm}{!}{$e_2$}};
        \vertex[below right =0.2cm and 0.8cm of 4](L2) {\resizebox{0.5cm}{!}{$e_4$}};

        \vertex[below right =0.8cm and 0.2cm of 2](L3) {\resizebox{0.5cm}{!}{$e_3$}};
        \vertex[below left =0.8cm and 0.2cm of 1](L4) {\resizebox{0.5cm}{!}{$e_1$}};

        \vertex[above left =0.05cm and 0.05cm of 1](LV1) {\resizebox{0.5cm}{!}{$v_1$}};
        \vertex[above right =0.05cm and 0.05cm of 2](LV2) {\resizebox{0.5cm}{!}{$v_2$}};
        \vertex[below right =0.05cm and 0.05cm of 3](LV3) {\resizebox{0.5cm}{!}{$v_3$}};
        \vertex[below left =0.05cm and 0.05cm of 4](LV4) {\resizebox{0.5cm}{!}{$v_4$}};

         \diagram*[large]{	
            (1) -- [->-,line width=0.6mm] (2),
            (2) -- [-<-,line width=0.6mm] (3),
            (3) -- [-<-,line width=0.6mm] (4),
            (4) -- [-<-,line width=0.6mm] (1),
        }; 
      \path[draw=black, fill=black] (1) circle[radius=0.05];
       \path[draw=black, fill=black] (2) circle[radius=0.05];
       \path[draw=black, fill=black] (3) circle[radius=0.05];
       \path[draw=black, fill=black] (4) circle[radius=0.05];
       
    \end{feynman}
    
\end{tikzpicture}
}}=\frac{i}{(\tilde{E}_1+\tilde{E}_2)}\left[\raisebox{-0.6cm}{\resizebox{1.6cm}{!}{
\begin{tikzpicture}

    \begin{feynman}
        \vertex(1);

        \vertex[right = 2cm of 1](2);

        \vertex[below = 2cm of 2](3);

        \vertex[left = 2cm of 3](4);

        \vertex[above right =0.8cm and 0.2cm of 3](L1) {\resizebox{0.5cm}{!}{$e_3$}};
        \vertex[below left =0.2cm and 0.8cm of 3](L2) {\resizebox{0.5cm}{!}{$e_4$}};

        \vertex[above right =1cm and 0.1cm of 4](L3) {\resizebox{0.5cm}{!}{$e_1$}};

        \vertex[above right =0.05cm and 0.05cm of 2](LV2) {\resizebox{0.7cm}{!}{$v_{12}$}};
        \vertex[below right =0.05cm and 0.05cm of 3](LV3) {\resizebox{0.5cm}{!}{$v_3$}};
        \vertex[below left =0.05cm and 0.05cm of 4](LV4) {\resizebox{0.5cm}{!}{$v_4$}};

         \diagram*[large]{	
            (2) -- [-<-,line width=0.6mm] (3),
            (3) -- [-<-,line width=0.6mm] (4),
            (4) -- [-<-,line width=0.6mm] (2),
        }; 
       \path[draw=black, fill=black] (2) circle[radius=0.05];
       \path[draw=black, fill=black] (3) circle[radius=0.05];
       \path[draw=black, fill=black] (4) circle[radius=0.05];
       
    \end{feynman}
    
\end{tikzpicture}
}}+\raisebox{-0.6cm}{\resizebox{1.6cm}{!}{
\begin{tikzpicture}

    \begin{feynman}
        \vertex(1);

        \vertex[right = 2cm of 1](2);

        \vertex[below = 2cm of 2](3);

        \vertex[left = 2cm of 3](4);

        \vertex[above right =0.8cm and 0.2cm of 3](L1) {\resizebox{0.5cm}{!}{$e_3$}};
        \vertex[below left =0.2cm and 0.8cm of 3](L2) {\resizebox{0.5cm}{!}{$e_4$}};

        \vertex[above right =1cm and 0.1cm of 4](L3) {\resizebox{0.5cm}{!}{$e_2$}};

        \vertex[above right =0.05cm and 0.05cm of 2](LV2) {\resizebox{0.5cm}{!}{$v_2$}};
        \vertex[below right =0.05cm and 0.05cm of 3](LV3) {\resizebox{0.5cm}{!}{$v_{3}$}};
        \vertex[below left =0.05cm and 0.05cm of 4](LV4) {\resizebox{0.7cm}{!}{$v_{14}$}};

         \diagram*[large]{	
            (2) -- [-<-,line width=0.6mm] (3),
            (3) -- [-<-,line width=0.6mm] (4),
            (4) -- [->-,line width=0.6mm] (2),
        }; 
       \path[draw=black, fill=black] (2) circle[radius=0.05];
       \path[draw=black, fill=black] (3) circle[radius=0.05];
       \path[draw=black, fill=black] (4) circle[radius=0.05];
       
    \end{feynman}
    
\end{tikzpicture}
}}\right].
\end{equation}
The first contracted graph we obtain has a directed cycle, and thus evaluates to zero. In other words, we can simply write:
\begin{equation}
\resizebox{2.cm}{!}{\raisebox{-1.5cm}{
\begin{tikzpicture}

    \begin{feynman}
        \vertex(1);

        \vertex[right = 2cm of 1](2);

        \vertex[below = 2cm of 2](3);

        \vertex[left = 2cm of 3](4);

        \vertex[above right =0.2cm and 0.8cm of 1](L1) {\resizebox{0.5cm}{!}{$e_2$}};
        \vertex[below right =0.2cm and 0.8cm of 4](L2) {\resizebox{0.5cm}{!}{$e_4$}};

        \vertex[below right =0.8cm and 0.2cm of 2](L3) {\resizebox{0.5cm}{!}{$e_3$}};
        \vertex[below left =0.8cm and 0.2cm of 1](L4) {\resizebox{0.5cm}{!}{$e_1$}};

        \vertex[above left =0.05cm and 0.05cm of 1](LV1) {\resizebox{0.5cm}{!}{$v_1$}};
        \vertex[above right =0.05cm and 0.05cm of 2](LV2) {\resizebox{0.5cm}{!}{$v_2$}};
        \vertex[below right =0.05cm and 0.05cm of 3](LV3) {\resizebox{0.5cm}{!}{$v_3$}};
        \vertex[below left =0.05cm and 0.05cm of 4](LV4) {\resizebox{0.5cm}{!}{$v_4$}};

         \diagram*[large]{	
            (1) -- [->-,line width=0.6mm] (2),
            (2) -- [-<-,line width=0.6mm] (3),
            (3) -- [-<-,line width=0.6mm] (4),
            (4) -- [-<-,line width=0.6mm] (1),
        }; 
      \path[draw=black, fill=black] (1) circle[radius=0.05];
       \path[draw=black, fill=black] (2) circle[radius=0.05];
       \path[draw=black, fill=black] (3) circle[radius=0.05];
       \path[draw=black, fill=black] (4) circle[radius=0.05];
       
    \end{feynman}
    
\end{tikzpicture}
}}=\frac{i}{(\tilde{E}_1+\tilde{E}_2)}\raisebox{-0.6cm}{\resizebox{1.6cm}{!}{
\begin{tikzpicture}

    \begin{feynman}
        \vertex(1);

        \vertex[right = 2cm of 1](2);

        \vertex[below = 2cm of 2](3);

        \vertex[left = 2cm of 3](4);

        \vertex[above right =0.8cm and 0.2cm of 3](L1) {\resizebox{0.5cm}{!}{$e_3$}};
        \vertex[below left =0.2cm and 0.8cm of 3](L2) {\resizebox{0.5cm}{!}{$e_4$}};

        \vertex[above right =1cm and 0.1cm of 4](L3) {\resizebox{0.5cm}{!}{$e_2$}};

        \vertex[above right =0.05cm and 0.05cm of 2](LV2) {\resizebox{0.5cm}{!}{$v_2$}};
        \vertex[below right =0.05cm and 0.05cm of 3](LV3) {\resizebox{0.5cm}{!}{$v_{3}$}};
        \vertex[below left =0.05cm and 0.05cm of 4](LV4) {\resizebox{0.7cm}{!}{$v_{14}$}};

         \diagram*[large]{	
            (2) -- [-<-,line width=0.6mm] (3),
            (3) -- [-<-,line width=0.6mm] (4),
            (4) -- [->-,line width=0.6mm] (2),
        }; 
       \path[draw=black, fill=black] (2) circle[radius=0.05];
       \path[draw=black, fill=black] (3) circle[radius=0.05];
       \path[draw=black, fill=black] (4) circle[radius=0.05];
       
    \end{feynman}
    
\end{tikzpicture}
}}.
\end{equation}
To iterate the contraction operation, we now choose $v_{14}$. We obtain
\begin{equation}
\resizebox{2.cm}{!}{\raisebox{-1.5cm}{
\begin{tikzpicture}

    \begin{feynman}
        \vertex(1);

        \vertex[right = 2cm of 1](2);

        \vertex[below = 2cm of 2](3);

        \vertex[left = 2cm of 3](4);

        \vertex[above right =0.2cm and 0.8cm of 1](L1) {\resizebox{0.5cm}{!}{$e_2$}};
        \vertex[below right =0.2cm and 0.8cm of 4](L2) {\resizebox{0.5cm}{!}{$e_4$}};

        \vertex[below right =0.8cm and 0.2cm of 2](L3) {\resizebox{0.5cm}{!}{$e_3$}};
        \vertex[below left =0.8cm and 0.2cm of 1](L4) {\resizebox{0.5cm}{!}{$e_1$}};

        \vertex[above left =0.05cm and 0.05cm of 1](LV1) {\resizebox{0.5cm}{!}{$v_1$}};
        \vertex[above right =0.05cm and 0.05cm of 2](LV2) {\resizebox{0.5cm}{!}{$v_2$}};
        \vertex[below right =0.05cm and 0.05cm of 3](LV3) {\resizebox{0.5cm}{!}{$v_3$}};
        \vertex[below left =0.05cm and 0.05cm of 4](LV4) {\resizebox{0.5cm}{!}{$v_4$}};

         \diagram*[large]{	
            (1) -- [->-,line width=0.6mm] (2),
            (2) -- [-<-,line width=0.6mm] (3),
            (3) -- [-<-,line width=0.6mm] (4),
            (4) -- [-<-,line width=0.6mm] (1),
        }; 
      \path[draw=black, fill=black] (1) circle[radius=0.05];
       \path[draw=black, fill=black] (2) circle[radius=0.05];
       \path[draw=black, fill=black] (3) circle[radius=0.05];
       \path[draw=black, fill=black] (4) circle[radius=0.05];
       
    \end{feynman}
    
\end{tikzpicture}
}}=\frac{i}{(\tilde{E}_1+\tilde{E}_2)}\frac{i}{(\tilde{E}_2+\tilde{E}_4)} \Bigg[\resizebox{1cm}{!}{\raisebox{-1.2cm}{
\begin{tikzpicture}

    \begin{feynman}
        \vertex(1);
        \vertex[below = 2cm of 1](3);

        \vertex[below left = 1cm and 0.3cm of 1](L1) {$e_4$};
        \vertex[below right = 1cm and 0.3cm of 1](L1) {$e_3$};

        \vertex[above =0.1cm of 1](LV1) {\resizebox{0.7cm}{!}{$v_{124}$}};
        \vertex[below =0.1cm of 3](LV2) {$v_3$};

         \diagram*[large]{	
            (1) -- [-<-,line width=0.6mm, out=-60, in=60] (3),
            (1) -- [->-,line width=0.6mm, out=240, in=120] (3),
        }; 
       \path[draw=black, fill=black] (1) circle[radius=0.05];
       \path[draw=black, fill=black] (3) circle[radius=0.05];
       
    \end{feynman}
    
\end{tikzpicture}
}}+\resizebox{1cm}{!}{\raisebox{-1.2cm}{
\begin{tikzpicture}

    \begin{feynman}
        \vertex(1);
        \vertex[below = 2cm of 1](3);

        \vertex[below left = 1cm and 0.3cm of 1](L1) {$e_2$};
        \vertex[below right = 1cm and 0.3cm of 1](L1) {$e_3$};

        \vertex[above =0.1cm of 1](LV1) {\resizebox{0.7cm}{!}{$v_{134}$}};
        \vertex[below =0.1cm of 3](LV2) {$v_{2}$};

         \diagram*[large]{	
            (1) -- [->-,line width=0.6mm, out=-60, in=60] (3),
            (1) -- [->-,line width=0.6mm, out=240, in=120] (3),
        }; 
       \path[draw=black, fill=black] (1) circle[radius=0.05];
       \path[draw=black, fill=black] (3) circle[radius=0.05];
       
    \end{feynman}
    
\end{tikzpicture}
}}\Bigg]=\frac{i}{(\tilde{E}_1+\tilde{E}_2)}\frac{i}{(\tilde{E}_2+\tilde{E}_4)} \resizebox{1cm}{!}{\raisebox{-1.2cm}{
\begin{tikzpicture}

    \begin{feynman}
        \vertex(1);
        \vertex[below = 2cm of 1](3);

        \vertex[below left = 1cm and 0.3cm of 1](L1) {$e_2$};
        \vertex[below right = 1cm and 0.3cm of 1](L1) {$e_3$};

        \vertex[above =0.1cm of 1](LV1) {\resizebox{0.7cm}{!}{$v_{134}$}};
        \vertex[below =0.1cm of 3](LV2) {$v_{2}$};

         \diagram*[large]{	
            (1) -- [->-,line width=0.6mm, out=-60, in=60] (3),
            (1) -- [->-,line width=0.6mm, out=240, in=120] (3),
        }; 
       \path[draw=black, fill=black] (1) circle[radius=0.05];
       \path[draw=black, fill=black] (3) circle[radius=0.05];
       
    \end{feynman}
    
\end{tikzpicture}
}}.
\end{equation}
To iterate one last time, we choose the vertex $v_2$ and contract all parallel edges ($e_2$ and $e_3$) at once, giving
\begin{equation}
\label{eq:or2_xfree}
\resizebox{2.cm}{!}{\raisebox{-1.5cm}{
\begin{tikzpicture}

    \begin{feynman}
        \vertex(1);

        \vertex[right = 2cm of 1](2);

        \vertex[below = 2cm of 2](3);

        \vertex[left = 2cm of 3](4);

        \vertex[above right =0.2cm and 0.8cm of 1](L1) {\resizebox{0.5cm}{!}{$e_2$}};
        \vertex[below right =0.2cm and 0.8cm of 4](L2) {\resizebox{0.5cm}{!}{$e_4$}};

        \vertex[below right =0.8cm and 0.2cm of 2](L3) {\resizebox{0.5cm}{!}{$e_3$}};
        \vertex[below left =0.8cm and 0.2cm of 1](L4) {\resizebox{0.5cm}{!}{$e_1$}};

        \vertex[above left =0.05cm and 0.05cm of 1](LV1) {\resizebox{0.5cm}{!}{$v_1$}};
        \vertex[above right =0.05cm and 0.05cm of 2](LV2) {\resizebox{0.5cm}{!}{$v_2$}};
        \vertex[below right =0.05cm and 0.05cm of 3](LV3) {\resizebox{0.5cm}{!}{$v_3$}};
        \vertex[below left =0.05cm and 0.05cm of 4](LV4) {\resizebox{0.5cm}{!}{$v_4$}};

         \diagram*[large]{	
            (1) -- [->-,line width=0.6mm] (2),
            (2) -- [-<-,line width=0.6mm] (3),
            (3) -- [-<-,line width=0.6mm] (4),
            (4) -- [-<-,line width=0.6mm] (1),
        }; 
      \path[draw=black, fill=black] (1) circle[radius=0.05];
       \path[draw=black, fill=black] (2) circle[radius=0.05];
       \path[draw=black, fill=black] (3) circle[radius=0.05];
       \path[draw=black, fill=black] (4) circle[radius=0.05];
       
    \end{feynman}
    
\end{tikzpicture}
}}
=\frac{i}{(\tilde{E}_1+\tilde{E}_2)}\frac{i}{(\tilde{E}_2+\tilde{E}_4)}\frac{i}{(\tilde{E}_2+\tilde{E}_3)}.
\end{equation}
This terminates the recursion.
\section{Cross-Free Families}

\noindent The choice of vertices that we performed at each iteration of the edge-contraction procedure is, in reality, not completely unconstrained, and engineered in order to obtain a result in a certain form. The specific set of rules that we are interested in, and that we followed in the examples above, is summarised in two constraints; at each iteration of the edge-contraction procedure, one must choose a vertex $v$ such that:
\begin{itemize}
\item $v$ is a sink or a source of the directed graph, meaning that all the arrows associated to the edges adjacent to it must either point towards $v$ or out of $v$. 
\item the complement of $v$, namely the subgraph with vertices given by $\mathcal{V}\setminus\{v\}$ and edges given by $\{\{v,v'\}\in\mathcal{E} \, | \, v,v'\in\vertices\setminus S\}$, must be connected.
\end{itemize}
It is relatively easy to check that the choices of the previous section indeed satisfy these two rules. For example, for the diagram
\begin{equation}
\resizebox{2.cm}{!}{\raisebox{-1.5cm}{
\begin{tikzpicture}

    \begin{feynman}
        \vertex(1);

        \vertex[right = 2cm of 1](2);

        \vertex[below = 2cm of 2](3);

        \vertex[left = 2cm of 3](4);

        \vertex[above right =0.2cm and 0.8cm of 1](L1) {\resizebox{0.5cm}{!}{$e_2$}};
        \vertex[below right =0.2cm and 0.8cm of 4](L2) {\resizebox{0.5cm}{!}{$e_4$}};

        \vertex[below right =0.8cm and 0.2cm of 2](L3) {\resizebox{0.5cm}{!}{$e_3$}};
        \vertex[below left =0.8cm and 0.2cm of 1](L4) {\resizebox{0.5cm}{!}{$e_1$}};

        \vertex[above left =0.05cm and 0.05cm of 1](LV1) {\resizebox{0.5cm}{!}{$v_1$}};
        \vertex[above right =0.05cm and 0.05cm of 2](LV2) {\resizebox{0.5cm}{!}{$v_2$}};
        \vertex[below right =0.05cm and 0.05cm of 3](LV3) {\resizebox{0.5cm}{!}{$v_3$}};
        \vertex[below left =0.05cm and 0.05cm of 4](LV4) {\resizebox{0.5cm}{!}{$v_4$}};

         \diagram*[large]{	
            (1) -- [->-,line width=0.6mm] (2),
            (2) -- [-<-,line width=0.6mm] (3),
            (3) -- [-<-,line width=0.6mm] (4),
            (4) -- [-<-,line width=0.6mm] (1),
        }; 
      \path[draw=black, fill=black] (1) circle[radius=0.05];
       \path[draw=black, fill=black] (2) circle[radius=0.05];
       \path[draw=black, fill=black] (3) circle[radius=0.05];
       \path[draw=black, fill=black] (4) circle[radius=0.05];
       
    \end{feynman}
    
\end{tikzpicture}
}},
\end{equation}
only the choices $v_1$ and $v_2$ satisfy the two constraints above. While all vertex choices satisfy the second constraint (for example, the complement of $v_1$, the graph $(\{v_2,v_3,v_4\},\{e_3,e_4\})$, is connected), $v_1$ and $v_2$ are the only sinks and sources of the graph. The existence, at each iteration of the edge-contraction procedure, of at least two vertices (at least one sink and at least one source) satisfying the properties above is guaranteed in general by the acyclic property of the graph. We will now see how this choice impacts the threshold singularity structure of the graph.

\subsection{Boundary operator}

\noindent For future use, let us introduce the boundary operator. For a set $S\subseteq \mathcal{V}$ of vertices, we have
\begin{equation}
\partial(S)=\{e=\{v,v'\}\in\mathcal{E} \, | \, v \in S, \ v'\notin S \ \ \text{or} \ \   v '\in S, \ v\notin S\}.
\end{equation}
For the labelling given in eq.~\eqref{eq:box_labelling} we have, for example, $\boundary(\{v_1\})=\{e_1,e_2\}$ (so that the boundary operator applied on a vertex simply gives the edges adjacent to that vertex) and $\boundary(\{v_1,v_2\})=\{e_1,e_3\}$ (in this case, the boundary operator gives all the edges adjacent to the vertex $v_{12}$ obtained by contracting $e_2$). We also define the characteristic vector, having one component for each edge of the graph:
\begin{equation}
\mathbf{1}^{\partial(S)}\in\{0,1\}^4, \quad \mathbf{1}^{\partial(S)}_i=\begin{cases}1 \quad &\text{if } e_i\in\boundary(S) \\ 0 \quad &\text{if } e_i\notin\boundary(S) \end{cases},
\end{equation}
so that, for example, $\mathbf{1}^{\partial(\{v_1\})}=(1,1,0,0)$. Cuts such as $S$ can be denoted graphically by circlings: for example
\begin{equation}
S=\{v_1,v_4\} \quad \Rightarrow \quad \resizebox{2.cm}{!}{\raisebox{-1.5cm}{
\begin{tikzpicture}

    \begin{feynman}
        \vertex(1);

        \vertex[right = 2cm of 1](2);

        \vertex[below = 2cm of 2](3);

        \vertex[below = 1cm of 1](M);

        \vertex[left = 2cm of 3](4);

        \vertex[above right =0.2cm and 0.8cm of 1](L1) {\resizebox{0.5cm}{!}{$e_2$}};
        \vertex[below right =0.2cm and 0.8cm of 4](L2) {\resizebox{0.5cm}{!}{$e_4$}};

        \vertex[below right =0.8cm and 0.2cm of 2](L3) {\resizebox{0.5cm}{!}{$e_3$}};
        \vertex[below left =0.8cm and 0.2cm of 1](L4) {\resizebox{0.5cm}{!}{$e_1$}};

        \vertex[above left =0.05cm and 0.05cm of 1](LV1) {\resizebox{0.5cm}{!}{$v_1$}};
        \vertex[above right =0.05cm and 0.05cm of 2](LV2) {\resizebox{0.5cm}{!}{$v_2$}};
        \vertex[below right =0.05cm and 0.05cm of 3](LV3) {\resizebox{0.5cm}{!}{$v_3$}};
        \vertex[below left =0.05cm and 0.05cm of 4](LV4) {\resizebox{0.5cm}{!}{$v_4$}};

         \diagram*[large]{	
            (1) -- [-,line width=0.6mm] (2),
            (2) -- [-,line width=0.6mm] (3),
            (3) -- [-,line width=0.6mm] (4),
            (4) -- [-,line width=0.6mm] (1),
        }; 
      \path[draw=black, fill=black] (1) circle[radius=0.05];
       \path[draw=black, fill=black] (2) circle[radius=0.05];
       \path[draw=black, fill=black] (3) circle[radius=0.05];
       \path[draw=black, fill=black] (4) circle[radius=0.05];

       \path[draw=red, line width=0.5mm] (M) ellipse (0.7cm and 1.3cm);
       
    \end{feynman}
    
\end{tikzpicture}
}}.
\end{equation}
The circling must contain in its interior the vertices of $S$, and the edges in $\boundary(S)$ are those crossed by the red line of the circling, namely $e_2$ and $e_4$.

\subsection{Decision trees}

\noindent Let us now interpret the result we obtained in the previous section by introducing an interesting graph-theoretic construct. Let us collect in a decision tree the choice of vertices that we made at each iteration, for the example orientation given in eq.~\eqref{eq:or1}. At the first iteration, we chose $v_1$ and got two terms; for one of them we chose $v_4$, while for the other we chose $v_2$; this, after eliminating all graphs that were not acyclic, also gave us a total of two terms; at the last iteration, we choose $v_3$ for both of them. In other words:
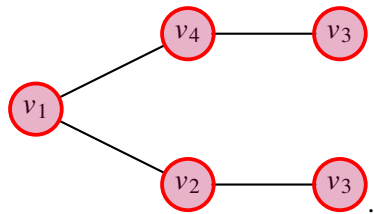
\begin{center}
\begin{tikzpicture}

    \begin{feynman}

        \vertex(2) {$v_1$}; 
        
        \vertex[above right =1cm and 2cm of 2](3) {$v_{4}$};
        \vertex[below right =1cm and 2cm of 2](4) {$v_{2}$};
        
        \vertex[right = 2cm of 3](5) {$v_{3}$}; 
        \vertex[right = 2cm of 4](6) {$v_{3}$};

         \diagram*[large]{

         (2) -- (3),
         (2) -- (4),
         (3) -- (5),
         (4) -- (6)

        }; 
      %\path[draw=red, fill=red!70!blue, opacity=0.3] (1) circle[radius=0.3];
       \path[draw=red, line width=0.5mm,fill=red!70!blue, fill opacity=0.3] (2) circle[radius=0.34];
       \path[draw=red, line width=0.5mm,fill=red!70!blue, fill opacity=0.3] (3) circle[radius=0.34];
       \path[draw=red, line width=0.5mm,fill=red!70!blue, fill opacity=0.3] (4) circle[radius=0.34];
       \path[draw=red, line width=0.5mm,fill=red!70!blue, fill opacity=0.3] (5) circle[radius=0.34];
       \path[draw=red, line width=0.5mm,fill=red!70!blue, fill opacity=0.3] (6) circle[radius=0.34];

       %\path[draw=black, fill=black] (C) circle[radius=0.05];
       
    \end{feynman}
    
\end{tikzpicture}.
\end{center}
The root of this tree is the starting choice of vertex, $v_1$, on the utmost left. The leaves of the tree are the last choices, $v_3$ and $v_3$, on the utmost right. We may now follow the unique path that connects each leaf to the root and collect the vertices that we encounter on the way. We find two families of subsets of vertices, corresponding to the two leaves of the tree:
\begin{equation}
F_1=\{\{v_1\},\{v_4\},\{v_3\}\}, \quad F_2=\{\{v_1\},\{v_2\},\{v_3\}\}.
\end{equation}
For the orientation given in eq.~\eqref{eq:or2}, the situation is different: there is only one term at the end of the recursion, and the choices of vertices are $v_1$, $v_{14}$ and $v_2$ or, in tree form
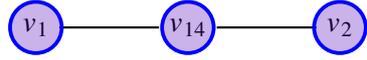
\begin{center}
\begin{tikzpicture}

    \begin{feynman}

        \vertex(2) {$v_1$}; 
        
        \vertex[right =2cm of 2](3) {$v_{14}$};
        \vertex[right = 2cm of 3](4) {$v_{2}$};

         \diagram*[large]{

         (2) -- (3),
         (3) -- (4),

        }; 
      %\path[draw=red, fill=red!70!blue, opacity=0.3] (1) circle[radius=0.3];
       \path[draw=blue, line width=0.5mm,fill=red!30!blue, fill opacity=0.3] (2) circle[radius=0.34];
       \path[draw=blue, line width=0.5mm,fill=red!30!blue, fill opacity=0.3] (3) circle[radius=0.34];
       \path[draw=blue, line width=0.5mm,fill=red!30!blue, fill opacity=0.3] (4) circle[radius=0.34];

       %\path[draw=black, fill=black] (C) circle[radius=0.05];
       
    \end{feynman}
    
\end{tikzpicture}.
\end{center}
The corresponding family of cuts is
\begin{equation}
F_3=\{\{v_1\},\{v_1,v_4\},\{v_2\}\},
\end{equation}
where the vertex $v_{14}$ is equivalent to the set $\{v_1,v_4\}$. These families can also be drawn using circlings: in particular, for $F_3$, we draw one circling for each cut that it contains, giving
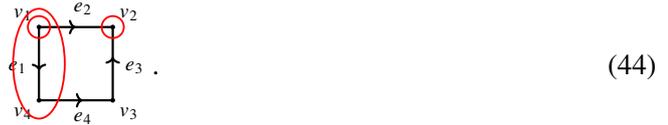
\begin{equation}
 \resizebox{2.cm}{!}{\raisebox{-1.5cm}{
\begin{tikzpicture}

    \begin{feynman}
        \vertex(1);

        \vertex[right = 2cm of 1](2);

        \vertex[below = 2cm of 2](3);

        \vertex[below = 1cm of 1](M);

        \vertex[left = 2cm of 3](4);

        \vertex[above right =0.2cm and 0.8cm of 1](L1) {\resizebox{0.5cm}{!}{$e_2$}};
        \vertex[below right =0.2cm and 0.8cm of 4](L2) {\resizebox{0.5cm}{!}{$e_4$}};

        \vertex[below right =0.8cm and 0.2cm of 2](L3) {\resizebox{0.5cm}{!}{$e_3$}};
        \vertex[below left =0.8cm and 0.2cm of 1](L4) {\resizebox{0.5cm}{!}{$e_1$}};

        \vertex[above left =0.05cm and 0.05cm of 1](LV1) {\resizebox{0.5cm}{!}{$v_1$}};
        \vertex[above right =0.05cm and 0.05cm of 2](LV2) {\resizebox{0.5cm}{!}{$v_2$}};
        \vertex[below right =0.05cm and 0.05cm of 3](LV3) {\resizebox{0.5cm}{!}{$v_3$}};
        \vertex[below left =0.05cm and 0.05cm of 4](LV4) {\resizebox{0.5cm}{!}{$v_4$}};

         \diagram*[large]{	
            (1) -- [->-,line width=0.6mm] (2),
            (2) -- [-<-,line width=0.6mm] (3),
            (3) -- [-<-,line width=0.6mm] (4),
            (4) -- [-<-,line width=0.6mm] (1),
        }; 
      \path[draw=black, fill=black] (1) circle[radius=0.05];
       \path[draw=black, fill=black] (2) circle[radius=0.05];
       \path[draw=black, fill=black] (3) circle[radius=0.05];
       \path[draw=black, fill=black] (4) circle[radius=0.05];

       \path[draw=red, line width=0.5mm] (M) ellipse (0.7cm and 1.5cm);

       \path[draw=red, line width=0.5mm] (1) ellipse (0.3cm and 0.3cm);

       \path[draw=red, line width=0.5mm] (2) ellipse (0.3cm and 0.3cm);
       
    \end{feynman}
    
\end{tikzpicture}
}}.
\end{equation}
For the family $F_1$, we have instead
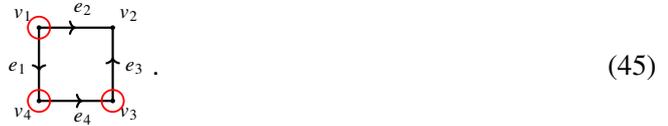
\begin{equation}
 \resizebox{2.cm}{!}{\raisebox{-1.5cm}{
\begin{tikzpicture}

    \begin{feynman}
        \vertex(1);

        \vertex[right = 2cm of 1](2);

        \vertex[below = 2cm of 2](3);

        \vertex[below = 1cm of 1](M);

        \vertex[left = 2cm of 3](4);

        \vertex[above right =0.2cm and 0.8cm of 1](L1) {\resizebox{0.5cm}{!}{$e_2$}};
        \vertex[below right =0.2cm and 0.8cm of 4](L2) {\resizebox{0.5cm}{!}{$e_4$}};

        \vertex[below right =0.8cm and 0.2cm of 2](L3) {\resizebox{0.5cm}{!}{$e_3$}};
        \vertex[below left =0.8cm and 0.2cm of 1](L4) {\resizebox{0.5cm}{!}{$e_1$}};

        \vertex[above left =0.05cm and 0.05cm of 1](LV1) {\resizebox{0.5cm}{!}{$v_1$}};
        \vertex[above right =0.05cm and 0.05cm of 2](LV2) {\resizebox{0.5cm}{!}{$v_2$}};
        \vertex[below right =0.05cm and 0.05cm of 3](LV3) {\resizebox{0.5cm}{!}{$v_3$}};
        \vertex[below left =0.05cm and 0.05cm of 4](LV4) {\resizebox{0.5cm}{!}{$v_4$}};

         \diagram*[large]{	
            (1) -- [->-,line width=0.6mm] (2),
            (2) -- [-<-,line width=0.6mm] (3),
            (3) -- [-<-,line width=0.6mm] (4),
            (4) -- [-<-,line width=0.6mm] (1),
        }; 
      \path[draw=black, fill=black] (1) circle[radius=0.05];
       \path[draw=black, fill=black] (2) circle[radius=0.05];
       \path[draw=black, fill=black] (3) circle[radius=0.05];
       \path[draw=black, fill=black] (4) circle[radius=0.05];

       \path[draw=red, line width=0.5mm] (1) ellipse (0.3cm and 0.3cm);

       \path[draw=red, line width=0.5mm] (3) ellipse (0.3cm and 0.3cm);

       \path[draw=red, line width=0.5mm] (4) ellipse (0.3cm and 0.3cm);
       
    \end{feynman}
    
\end{tikzpicture}
}}.
\end{equation}
A look at these diagrams makes it clear that the families we are investigating are constrained heavily.

\subsection{Threshold singularity structure}

\noindent We first observe that eq.~\eqref{eq:end_contraction} can be rewritten as
\begin{equation}
\resizebox{2.cm}{!}{\raisebox{-1.5cm}{
\begin{tikzpicture}

    \begin{feynman}
        \vertex(1);

        \vertex[right = 2cm of 1](2);

        \vertex[below = 2cm of 2](3);

        \vertex[left = 2cm of 3](4);

        \vertex[above right =0.2cm and 0.8cm of 1](L1) {\resizebox{0.5cm}{!}{$e_2$}};
        \vertex[below right =0.2cm and 0.8cm of 4](L2) {\resizebox{0.5cm}{!}{$e_4$}};

        \vertex[below right =0.8cm and 0.2cm of 2](L3) {\resizebox{0.5cm}{!}{$e_3$}};
        \vertex[below left =0.8cm and 0.2cm of 1](L4) {\resizebox{0.5cm}{!}{$e_1$}};

        \vertex[above left =0.05cm and 0.05cm of 1](LV1) {\resizebox{0.5cm}{!}{$v_1$}};
        \vertex[above right =0.05cm and 0.05cm of 2](LV2) {\resizebox{0.5cm}{!}{$v_2$}};
        \vertex[below right =0.05cm and 0.05cm of 3](LV3) {\resizebox{0.5cm}{!}{$v_3$}};
        \vertex[below left =0.05cm and 0.05cm of 4](LV4) {\resizebox{0.5cm}{!}{$v_4$}};

         \diagram*[large]{	
            (1) -- [->-,line width=0.6mm] (2),
            (2) -- [-<-,line width=0.6mm] (3),
            (3) -- [->-,line width=0.6mm] (4),
            (4) -- [-<-,line width=0.6mm] (1),
        }; 
      \path[draw=black, fill=black] (1) circle[radius=0.05];
       \path[draw=black, fill=black] (2) circle[radius=0.05];
       \path[draw=black, fill=black] (3) circle[radius=0.05];
       \path[draw=black, fill=black] (4) circle[radius=0.05];
       
    \end{feynman}
    
\end{tikzpicture}
}}=\sum_{j=1}^2 \prod_{S\in F_j} \frac{i}{ \mathbf{1}^{\boundary(S)} \cdot \tilde{\mathbf{E}}^{\vec{\sigma}_3}},
\end{equation}
while eq.~\eqref{eq:or2_xfree} can be written as
\begin{equation}
\resizebox{2.cm}{!}{\raisebox{-1.5cm}{
\begin{tikzpicture}

    \begin{feynman}
        \vertex(1);

        \vertex[right = 2cm of 1](2);

        \vertex[below = 2cm of 2](3);

        \vertex[left = 2cm of 3](4);

        \vertex[above right =0.2cm and 0.8cm of 1](L1) {\resizebox{0.5cm}{!}{$e_2$}};
        \vertex[below right =0.2cm and 0.8cm of 4](L2) {\resizebox{0.5cm}{!}{$e_4$}};

        \vertex[below right =0.8cm and 0.2cm of 2](L3) {\resizebox{0.5cm}{!}{$e_3$}};
        \vertex[below left =0.8cm and 0.2cm of 1](L4) {\resizebox{0.5cm}{!}{$e_1$}};

        \vertex[above left =0.05cm and 0.05cm of 1](LV1) {\resizebox{0.5cm}{!}{$v_1$}};
        \vertex[above right =0.05cm and 0.05cm of 2](LV2) {\resizebox{0.5cm}{!}{$v_2$}};
        \vertex[below right =0.05cm and 0.05cm of 3](LV3) {\resizebox{0.5cm}{!}{$v_3$}};
        \vertex[below left =0.05cm and 0.05cm of 4](LV4) {\resizebox{0.5cm}{!}{$v_4$}};

         \diagram*[large]{	
            (1) -- [->-,line width=0.6mm] (2),
            (2) -- [-<-,line width=0.6mm] (3),
            (3) -- [-<-,line width=0.6mm] (4),
            (4) -- [-<-,line width=0.6mm] (1),
        }; 
      \path[draw=black, fill=black] (1) circle[radius=0.05];
       \path[draw=black, fill=black] (2) circle[radius=0.05];
       \path[draw=black, fill=black] (3) circle[radius=0.05];
       \path[draw=black, fill=black] (4) circle[radius=0.05];
       
    \end{feynman}
    
\end{tikzpicture}
}}=\prod_{S\in F_3} \frac{i}{ \mathbf{1}^{\boundary(S)} \cdot \tilde{\mathbf{E}}^{\vec{\sigma}_4}},
\end{equation}
where we introduced the symbol $\tilde{\mathbf{E}}^{\vec{\sigma}}$ such that $(\tilde{\mathbf{E}}^{\vec{\sigma}})_i=E_i-\vec{\sigma}_i p_i^0$. Furthermore, the families $F_1$, $F_2$ and $F_3$ satisfy striking properties:
\begin{itemize}
\item For $S\in F_i$,
%, $S$ and $\mathcal{V}\setminus S$ identify two subgraphs, $(S,\{\{v,v'\}\in\mathcal{E}\, | \, v,v'\in S\})$ and $(\vertices\setminus S,\{\{v,v'\}\in\mathcal{E}\, | \, v,v'\in \vertices\setminus S\})$), namely 
each of the two subgraphs obtained from the original graph by deleting the edges in $\boundary(S)$ is \emph{connected}.
\item For two distinct elements of the family $S,S'\in F_i$, one of the following must hold: $S\subset S'$ or $S'\subset S$ or $S\cap S'=\emptyset$.
\item Any set of $F_i$ cannot be written as a union of other sets in $F_i$.
\end{itemize}
Let us verify these properties for the family $F_1$. We have, indeed, that $S=\{v_1\}\in F_1$ identifies a connected subgraph (the graph made up by just one vertex) and $\mathcal{V}\setminus S=\{v_2,v_3,v_4\}$ also identifies a connected subgraph (namely, the subgraph $(\mathcal{V}\setminus S, \{e_3,e_4\})$). An analogous argument holds for all other cuts in $F_1$. Clearly, the second property is also satisfied, since in this case for any $S,S'\in F_1$, we have $S\cap S'=\emptyset$. The third property is also trivial to check.

\noindent The analysis we performed can be extended to all of the acyclic orientations of the box. In summary, one obtains
\begin{equation}
\label{eq:x_free}
\threerep_\Box=\sum_{\substack{\vec{\sigma} \text{ s.t.} \\ G_{\vec{\sigma}} \text{ is acyclic}}} \frac{\mathcal{N}(\sigma_1E_1,\sigma_2E_2,\sigma_3E_3,\sigma_4E_4)}{\prod_{j=1}^4 2i E_j} \sum_{F\in\mathcal{F}_{\vec{\sigma}}} \prod_{S\in F} \frac{i}{ \mathbf{1}^{\boundary(S)} \cdot \tilde{\mathbf{E}}^{\vec{\sigma}}},
\end{equation}
where $\mathcal{F}_{\vec{\sigma}}$ collects the families obtained by applying the contraction operation to the orientation $\vec{\sigma}$ (for example, for $\vec{\sigma}=(-1,1,-1,1)$, we have $\mathcal{F}_{\vec{\sigma}}=\{F_1,F_2\}$ and for $\vec{\sigma}=(-1,1,-1,-1)$, we have $\mathcal{F}_{\vec{\sigma}}=\{F_3\}$). Eq.~\eqref{eq:x_free} gives the Cross-Free Family representation for the box diagram.

\section{Comparison with Time-Ordered Perturbation Theory}

\noindent The Cross-Free Family representation and the Time-Ordered Perturbation Theory representation (see ref.~\cite{tasi_sterman,Sterman:1993hfp} for details on the TOPT representation) are related by partial fractioning relations. For the box example, they are simple enough that they can be worked out explicitly. In general, performing such partial fractioning procedure for more complex topologies constitutes a complicated combinatorical problem. Starting from eq.~\eqref{eq:end_contraction}, for example, we may write
\begin{equation}
\label{eq:topt}
\resizebox{2.cm}{!}{\raisebox{-1.5cm}{
\begin{tikzpicture}

    \begin{feynman}
        \vertex(1);

        \vertex[right = 2cm of 1](2);

        \vertex[below = 2cm of 2](3);

        \vertex[left = 2cm of 3](4);

        \vertex[above right =0.2cm and 0.8cm of 1](L1) {\resizebox{0.5cm}{!}{$e_2$}};
        \vertex[below right =0.2cm and 0.8cm of 4](L2) {\resizebox{0.5cm}{!}{$e_4$}};

        \vertex[below right =0.8cm and 0.2cm of 2](L3) {\resizebox{0.5cm}{!}{$e_3$}};
        \vertex[below left =0.8cm and 0.2cm of 1](L4) {\resizebox{0.5cm}{!}{$e_1$}};

        \vertex[above left =0.05cm and 0.05cm of 1](LV1) {\resizebox{0.5cm}{!}{$v_1$}};
        \vertex[above right =0.05cm and 0.05cm of 2](LV2) {\resizebox{0.5cm}{!}{$v_2$}};
        \vertex[below right =0.05cm and 0.05cm of 3](LV3) {\resizebox{0.5cm}{!}{$v_3$}};
        \vertex[below left =0.05cm and 0.05cm of 4](LV4) {\resizebox{0.5cm}{!}{$v_4$}};

         \diagram*[large]{	
            (1) -- [->-,line width=0.6mm] (2),
            (2) -- [-<-,line width=0.6mm] (3),
            (3) -- [->-,line width=0.6mm] (4),
            (4) -- [-<-,line width=0.6mm] (1),
        }; 
      \path[draw=black, fill=black] (1) circle[radius=0.05];
       \path[draw=black, fill=black] (2) circle[radius=0.05];
       \path[draw=black, fill=black] (3) circle[radius=0.05];
       \path[draw=black, fill=black] (4) circle[radius=0.05];
       
    \end{feynman}
    
\end{tikzpicture}
}}=\frac{i}{(\tilde{E}_1+\tilde{E}_2+\tilde{E}_3+\tilde{E}_4)}\Bigg[\frac{i^2}{(\tilde{E}_2+\tilde{E}_4)(\tilde{E}_3+\tilde{E}_4)}+\frac{i^2}{(\tilde{E}_2+\tilde{E}_4)(\tilde{E}_1+\tilde{E}_2)} \nonumber
\end{equation}
\begin{equation}+\frac{i^2}{(\tilde{E}_1+\tilde{E}_4)(\tilde{E}_3+\tilde{E}_4)}+\frac{i^2}{(\tilde{E}_1+\tilde{E}_4)(\tilde{E}_1+\tilde{E}_2)}\Bigg].
\end{equation}
Each of the four terms on the right-hand side is a term appearing in the Time-Ordered Perturbation Theory representation. In particular, we may write the diagrammatic identity:
\begin{equation}
\resizebox{2.cm}{!}{\raisebox{-1.5cm}{
\begin{tikzpicture}

    \begin{feynman}
        \vertex(1);

        \vertex[right = 2cm of 1](2);

        \vertex[below = 2cm of 2](3);

        \vertex[below = 1cm of 1](M);

        \vertex[left = 2cm of 3](4);

        \vertex[above right =0.2cm and 0.8cm of 1](L1) {\resizebox{0.5cm}{!}{$e_2$}};
        \vertex[below right =0.2cm and 0.8cm of 4](L2) {\resizebox{0.5cm}{!}{$e_4$}};

        \vertex[below right =0.8cm and 0.2cm of 2](L3) {\resizebox{0.5cm}{!}{$e_3$}};
        \vertex[below left =0.8cm and 0.2cm of 1](L4) {\resizebox{0.5cm}{!}{$e_1$}};

        \vertex[above left =0.05cm and 0.05cm of 1](LV1) {\resizebox{0.5cm}{!}{$v_1$}};
        \vertex[above right =0.05cm and 0.05cm of 2](LV2) {\resizebox{0.5cm}{!}{$v_2$}};
        \vertex[below right =0.05cm and 0.05cm of 3](LV3) {\resizebox{0.5cm}{!}{$v_3$}};
        \vertex[below left =0.05cm and 0.05cm of 4](LV4) {\resizebox{0.5cm}{!}{$v_4$}};

         \diagram*[large]{	
            (1) -- [->-,line width=0.6mm] (2),
            (2) -- [-<-,line width=0.6mm] (3),
            (3) -- [-<-,line width=0.6mm] (4),
            (4) -- [-<-,line width=0.6mm] (1),
        }; 
      \path[draw=black, fill=black] (1) circle[radius=0.05];
       \path[draw=black, fill=black] (2) circle[radius=0.05];
       \path[draw=black, fill=black] (3) circle[radius=0.05];
       \path[draw=black, fill=black] (4) circle[radius=0.05];

       \path[draw=red, line width=0.5mm] (1) ellipse (0.3cm and 0.3cm);

       \path[draw=red, line width=0.5mm] (3) ellipse (0.3cm and 0.3cm);

       \path[draw=red, line width=0.5mm] (4) ellipse (0.3cm and 0.3cm);
       
    \end{feynman}
    
\end{tikzpicture}
}}+\resizebox{2.cm}{!}{\raisebox{-1.5cm}{
\begin{tikzpicture}

    \begin{feynman}
        \vertex(1);

        \vertex[right = 2cm of 1](2);

        \vertex[below = 2cm of 2](3);

        \vertex[below = 1cm of 1](M);

        \vertex[left = 2cm of 3](4);

        \vertex[above right =0.2cm and 0.8cm of 1](L1) {\resizebox{0.5cm}{!}{$e_2$}};
        \vertex[below right =0.2cm and 0.8cm of 4](L2) {\resizebox{0.5cm}{!}{$e_4$}};

        \vertex[below right =0.8cm and 0.2cm of 2](L3) {\resizebox{0.5cm}{!}{$e_3$}};
        \vertex[below left =0.8cm and 0.2cm of 1](L4) {\resizebox{0.5cm}{!}{$e_1$}};

        \vertex[above left =0.05cm and 0.05cm of 1](LV1) {\resizebox{0.5cm}{!}{$v_1$}};
        \vertex[above right =0.05cm and 0.05cm of 2](LV2) {\resizebox{0.5cm}{!}{$v_2$}};
        \vertex[below right =0.05cm and 0.05cm of 3](LV3) {\resizebox{0.5cm}{!}{$v_3$}};
        \vertex[below left =0.05cm and 0.05cm of 4](LV4) {\resizebox{0.5cm}{!}{$v_4$}};

         \diagram*[large]{	
            (1) -- [->-,line width=0.6mm] (2),
            (2) -- [-<-,line width=0.6mm] (3),
            (3) -- [-<-,line width=0.6mm] (4),
            (4) -- [-<-,line width=0.6mm] (1),
        }; 
      \path[draw=black, fill=black] (1) circle[radius=0.05];
       \path[draw=black, fill=black] (2) circle[radius=0.05];
       \path[draw=black, fill=black] (3) circle[radius=0.05];
       \path[draw=black, fill=black] (4) circle[radius=0.05];

       \path[draw=red, line width=0.5mm] (1) ellipse (0.3cm and 0.3cm);

       \path[draw=red, line width=0.5mm] (3) ellipse (0.3cm and 0.3cm);

       \path[draw=red, line width=0.5mm] (2) ellipse (0.3cm and 0.3cm);
       
    \end{feynman}
    
\end{tikzpicture}
}}=\raisebox{-0.6cm}{\includegraphics[page=62]{tikz.pdf}}\hspace{4cm}\nonumber\end{equation}
\begin{equation}\hspace{1cm}+\raisebox{-0.6cm}{\includegraphics[page=63]{tikz.pdf}}+\raisebox{-0.6cm}{\includegraphics[page=64]{tikz.pdf}}+\raisebox{-0.6cm}{\includegraphics[page=65]{tikz.pdf}}.
\end{equation}
What is the convenience of the Cross-Free Family representation, then? The denominator, appearing in eq.~\eqref{eq:topt}, in the form $(\tilde{E}_1+\tilde{E}_2+\tilde{E}_3+\tilde{E}_4)$, does not appear in the Cross-Free Family representation. In other words, it corresponds to a \emph{spurious singularity} of the Time-Ordered Perturbation Theory representation. Such denominator corresponds to a cut, the middle one appearing in all the TOPT diagrams and that cuts all edges of the graph, that divides the graph in four connected components (the four singleton vertices obtained by deleting the edges crossed by the cut). This rule is indeed in contradiction with that characterising the three families $F_1$, $F_2$ and $F_3$ from the previous section.

\section{Conclusion}
\noindent We have shown how to derive the Cross-Free Family representation for the box diagrams and used it to constrain its singularity structure. The Cross-Free Family representation has numerical advantages, in that it is a) relatively compact when compared to that of ref.~\cite{capatti2020manifestly}, especially when numerators are involved, b) numerically stable, since it does not have spurious singularities. The Cross-Free Family representation also has numerous theoretical advantages: the absence of spurious singularities primes it to be a central tool in the singularity analysis of Feynman diagrams and related proofs (such as that of ref.~\cite{Capatti:2020xjc}). In the future, it would be interesting to study its consequences on the analytic structure of Feynman diagrams (see, for example, the use of Time Ordered Perturbation Theory~\cite{bourjaily2020sequential} to this end). The privileged role of the principle of connectedness in the Cross-Free Family representation unveils the diagrammatic manifestation of the cluster decomposition principle: for this reason, it would be interesting to study an operator-level derivation of the Cross-Free Family representation, analogous to that yielding the TOPT representation~\cite{schwartz2014quantum}.

\bibliography{biblio}
\bibliographystyle{unsrt}

\end{document}